\documentclass[a4paper,11pt]{article}
\usepackage{subfig}
\usepackage{jheppub}
\usepackage{multirow}
\usepackage{amsmath}
\usepackage{hyperref}
\usepackage{graphicx}

\pdfoutput=1

\let\jnfont=\rm
\def\NPB#1,{{\jnfont Nucl.\ Phys.\ B }{\bf #1},}
\def\PLB#1,{{\jnfont Phys.\ Lett.\ B }{\bf #1},}
\def\EPJC#1,{{\jnfont Eur.\ Phys.\ Jour.\ C }{\bf #1},}
\def\PRD#1,{{\jnfont Phys.\ Rev.\ D }{\bf #1},}
\def\PRL#1,{{\jnfont Phys.\ Rev.\ Lett.\ }{\bf #1},}
\def\MPLA#1,{{\jnfont Mod.\ Phys.\ Lett.\ A }{\bf #1},}
\def\JPG#1,{{\jnfont J.\ Phys.\ G}{\bf #1},}
\def\CTP#1,{{\jnfont Commun.\ Theor.\ Phys.\ }{\bf #1},}
\def\ZPC#1,{{\jnfont Z.\ Phys.\ C }{\bf #1},}
\def\JHEP#1,{{\jnfont JHEP \ }{\bf #1},}

\title{Sneutrino DM in the NMSSM with inverse seesaw mechanism}

\author{Junjie Cao$^{1,2,3}$, Xiaofei Guo$^1$, Yangle He$^1$, Liangliang Shang$^1$, Yuanfang Yue$^1$}

\affiliation{ $^1$  College of Physics and Materials Science,
        Henan Normal University, Xinxiang 453007, China \\
 $^2$ Center for High Energy Physics, Peking University, Beijing 100871, China \\
 $^3$ Department of Applied Physics, Xi'an Jiaotong University, Xi'an 710049, China \\ }

\emailAdd{junjiec@itp.ac.cn}
\emailAdd{guoxf@gs.zzu.edu.cn}
\emailAdd{heyangle90@gmail.com}
\emailAdd{shlwell1988@gmail.com}
\emailAdd{yuanfang405@gmail.com}

\abstract{In supersymmetric theories like the Next-to-Minimal Supersymmetric Standard Model (NMSSM), the lightest neutralino with bino or singlino as its dominant component is customarily taken as dark matter (DM) candidate. Since light Higgsinos favored by naturalness can strength the couplings of the DM and thus enhance the DM-nucleon scattering rate, the tension between naturalness and DM direct detection results becomes more and more acute with the improved experimental sensitivity. In this work, we extend the NMSSM by inverse seesaw mechanism to generate neutrino mass, and show that in certain parameter space the lightest sneutrino may act as a viable DM candidate, i.e. it can annihilate by multi-channels to get correct relic density and meanwhile satisfy all experimental constraints. The most striking feature of the extension is that the DM-nucleon scattering rate can be naturally below its current experimental bounds regardless of the higgsino mass, and hence it alleviates the tension between naturalness and DM experiments. Other interesting features  include that the Higgs phenomenology becomes much richer than that of the original NMSSM due to the relaxed constraints from DM physics and also due to the presence of extra neutrinos, and that the signatures of sparticles at colliders are quite different from those with neutralino as DM candidate.}

\begin{document}
\maketitle \indent
\newpage

\section{\label{introduction}Introduction}

From recent cosmological and astrophysical measurements with unprecedented precision, it has been a robust fact that over $20\%$ of the energy density of the Universe today
is composed of Dark Matter (DM) \cite{Ade:2015xua}. Among various kinds of DM candidates, the massive neutral stable particle with weak couplings to quarks/leptons is a promising one, and has been widely discussed in different new physics models for past decades. In the popular
supersymmetric models such as the Minimal Supersymmetric Standard Model (MSSM) \cite{Haber:1984rc,Gunion:1984yn}, the lightest neutralino
with bino field as its dominant component has such properties \cite{Goldberg:1983nd,Ellis:1983ew,Jungman:1995df},
and is customarily treated as DM candidate in phenomenological study. In this setup,
the interactions of the DM with Higgs bosons are inversely proportional to higgsino mass $\mu$ \cite{Cao:2015efs}, and
the lighter the higgsino is, the stronger the couplings become.
This in return results in an increased DM-nucleon scattering rate \footnote{We emphasize here that we only consider the case of one-component DM
with its mass at electro-weak scale. In this case, the lightest neutralino in the MSSM is the admixture of gaugino and higgsino with
bino as its largest component field in order to predict the right relic DM density. Alternatively if the lightest neutralino is an almost
pure higgsino which can be realized in natural SUSY \cite{Baer:2012uy} or mirage mediation scenarios \cite{Allahverdi:2012wb},
its current density will fall far short to account for the measured value of DM density
\cite{Baer:2012uy,Allahverdi:2012wb}, and meanwhile the corresponding DM-nucleon scattering rate is usually suppressed
too \cite{Perelstein:2012qg}. Note that the tendency of a light $\mu$ to enhance DM-nucleon scattering rate is also
applied to the NMSSM where the lightest neutralino is
usually bino-dominated or singlino-dominated \cite{Cao:2015loa}. }. On the other hand, the higgsino
mass determines at tree level the $Z$ boson mass, and naturalness favors light higgsinos up to several hundred ${\rm GeV}$ \cite{Baer:2012uy}.
Obviously, with the rapidly improved sensitivity of DM direct
detection (DD) experiments such as PandaX-II \cite{Tan:2016zwf,Fu:2016ega}, LUX \cite{Akerib:2016vxi} and XENON-1T \cite{Aprile:2017iyp} to
DM-nucleon scattering rate in recent years, there emerges increasing tension between naturalness and the DD experiments \cite{Baer:2016ucr,Cao:2016cnv}.
Confronted with such a situation, some authors recently emphasized the role of blind spots in escaping the strong constraints from the DD experiments
\cite{Crivellin:2015bva,Badziak:2015exr,Han:2016qtc,Badziak:2017uto}.
These parameter points, however, require subtle cancelation among different contributions to the scattering rate, and hence lead to a certain degree of
fine tuning. Another long-standing problem that the MSSM fails to account for comes from the observation of neutrino oscillation, which can be explained only
if neutrinos have tiny masses \cite{Mohapatra:2005wg,Strumia:2006db,King:2013eh}. Given the fact that the MSSM with R-parity conservation has no built-in mechanism to generate the masses, neutrino
oscillation indicates unambiguously the existence of extra physics.

In this work, we intend to seek for the theory that can address the origin of neutrino mass and the nature of DM simultaneously.
To be more specific, we require it to have following properties:
\begin{itemize}
\item predicting in a natural way the masses of active neutrinos and also the recently discovered Higgs boson
      with its field content as economical as possible;
\item providing a testable mechanism to generate sterile neutrino masses;
\item easily satisfying the experimental data such as the neutrino oscillation data, the electroweak precision measurements, and the lepton-flavor violation;
\item easily coinciding with the observations in DM physics even for light higgsinos, especially that DM-nucleon scattering rate should be naturally
suppressed to satisfy the very tight constraints from the recent XENON-1T experiment.
\end{itemize}
In constructing such a theory, we note that among the ideas to generate the tiny neutrino masses, the inverse seesaw mechanism \cite{Mohapatra:1986bd} is rather attractive since
in its configuration, the smallness of neutrino masses is attributed to lepton number violation (LNV) and a doubly suppressed ratio, all involved dimensional parameters are at weak scale and the
Yukawa couplings of the neutrinos may be moderately large, all of which indicate that the mechanism can provide a natural, simple and
testable way to realize the small neutrino masses at low energy \cite{Xing:2009in}.
We also note that the gauge singlet extensions of the MSSM like the Next-to-Minimal
Supersymmetric Standard Model (NMSSM) \cite{Ellwanger:2009dp} have great theoretical advantages, e.g. their capability of generating dynamically the higgsino mass $\mu$ and enhancing the
SM-like Higgs boson mass by the singlet-doublet interaction among the Higgs fields in the theory and/or by the singlet-doublet Higgs mixing effect \cite{Ellwanger:2011aa,Cao:2012fz}.
These features motivate us to incorporate the inverse seesaw mechanism in the NMSSM as an attempt at weak scale to solve the problems mentioned above. Interestingly, we find that the
resulting theory not only inherits all the merits of the NMSSM and
the seesaw mechanism, but also exhibits following new features:
\begin{itemize}
\item Except for the tiny Majorana masses for extra family of sterile neutrino fields, which violates lepton number by two units and
is naturally small according to 't Hooft's naturalness criterion \cite{Hooft}, there is no dimensional parameters in its superpotential.
As a result, the mass for any new particle beyond the Standard Model, such as sterile neutrinos and supersymmetric particles,
is determined by the vacuum expectation values (vev) of Higgs fields and/or by soft supersymmetry breaking coefficients .
\item The lightest sneutrino $\tilde{\nu}_1$ may act as a viable DM candidate. In more detail, unlike some pioneer studies
in this direction \cite{sneutrino-1,sneutrino-2,sneutrino-3,sneutrino-4,sneutrino-5,
sneutrino-6,sneutrino-7,sneutrino-8,sneutrino-9,sneutrino-10}, the sneutrino DM in our framework has two attractive characters.
One is that $\tilde{\nu}_1$ can mainly annihilate into a pair of singlet dominated Higgs bosons
to get the right relic density and meanwhile satisfy all experimental constraints. This process is determined by the interactions
of $\tilde{\nu}_1$ with the singlet
Higgs fields for a given Higgs boson spectrum, and consequently DM observables are sensitive only to the parameters in sneutrino sector. The other is owe to the fact that
the singlet field can mediate the transition between $\tilde{\nu}_1$ pair and the higgsino pair, which implies
that  $\tilde{\nu}_1$ and the higgsinos can be in thermal equilibrium in early Universe before their freeze-out. If their mass splitting is less than about $10\%$,
the number density of the higgsinos can track that of  $\tilde{\nu}_1$ during freeze-out, and consequently the higgsinos played an important role in determining DM relic density
\cite{Coannihilation} (in literature such a phenomenon was called coannihilation \cite{Griest:1990kh}). As a result, even for very weak  couplings of
$\tilde{\nu}_1$ with SM particles, $\tilde{\nu}_1$
may still reach the correct relic density by coannihilating with the higgsino-dominated particles.  Again, this translates to the constraints only on the parameters
in sneutrino sector if the higgsino mass is less than the other neutralino masses.

Due to the mentioned properties of $\tilde{\nu}_1$, the DM-nucleon scattering rate in our model can be naturally suppressed by the small mixing between singlet-doublet
Higgs fields (corresponding to the former case) or by the highly sterile nature of $\tilde{\nu}_1$ (the latter case).  This suppression is
independent of the parameter $\mu$, and hence there is no tension any more between the weak scale naturalness
and DM physics.

\item Due to potentially relaxed DM constraints on the theory and also due to the presence of possible light sterile neutrinos, Higgs physics is enriched greatly compared with that of
the unextended NMSSM. Moreover, the signature of sparticles at colliders is greatly changed for sneutrino DM instead of the customary neutralino DM.
\end{itemize}

With respect to these features, we have more explanations. One is that in the original MSSM and NMSSM, only left-handed sneutrinos are
predicted, and consequently the sneutrino $\tilde{\nu}_1$ as DM candidate is excluded by DD experiments due to its sizable coupling with $Z$ boson \cite{Arina:2007tm}.
In Type-I seesaw extended models, however, sneutrino may be a viable DM because the inclusion of right-handed (RH) neutrino superfields
in the theory enables the DM to be RH sneutrino dominated, which can reduce the coupling strength greatly  \cite{NMSSM-SS-1}.
In the inverse seesaw extension, beside the RH fields an extra family of sterile neutrino fields are also introduced, which is able to further suppress
the left-handed sneutrino component of $\tilde{\nu}_1$ to get a smaller DM-nucleon scattering rate. In fact, this is one of the reasons that we are interested in the incorporation of the inverse seesaw mechanism within supersymmetric theories. The other is that the features mentioned above are not unique to
the inverse seesaw extension of the NMSSM. In fact, the Type-I seesaw extension of the NMSSM also possesses these properties, and in particular it
has an advantage over our framework in that it corresponds to a more economical field assignment  \cite{NMSSM-SS-1}. However,  as we will discuss at the end of this work,
our framework provides more flexibility to accommodate low energy data (such as the neutrino oscillation data, the electroweak precision measurements and the lepton-flavor violation) and richer phenomenology than the Type-I seesaw extension, which make it
worthy of an intensive study.

The main purpose of this work is to illustrate the properties of the sneutrino DM in the NMSSM with inverse seesaw mechanism (ISS-NMSSM). For this end, we vary the
parameters in sneutrino sector to obtain physical parameter points, and show how $\tilde{\nu}_1$ annihilated to get the right relic density and meanwhile
avoids the constraints from Fermi-LAT search for DM annihilation
in dwarf galaxies. In particular, we pay great attention to study DM-nucleon scattering, and exhibit suppression mechanisms of the theory on the rate.
We note that the inverse seesaw mechanism has been intensively studied in the framework of the MSSM \cite{MSSM-ISS-1,MSSM-ISS-2,MSSM-ISS-3,MSSM-ISS-4,
MSSM-ISS-5,MSSM-ISS-6,MSSM-ISS-7,MSSM-ISS-8,MSSM-ISS-9,MSSM-ISS-10,MSSM-ISS-11,MSSM-ISS-12,MSSM-ISS-13,MSSM-ISS-14,MSSM-ISS-15,MSSM-ISS-16,
MSSM-ISS-17,MSSM-ISS-18,MSSM-ISS-19,MSSM-ISS-20}
and the supersymmetric B-L models \cite{B-L-ISS-1,B-L-ISS-2,B-L-ISS-3,B-L-ISS-4,B-L-ISS-5,B-L-ISS-6,B-L-ISS-7,B-L-ISS-8,B-L-ISS-9,B-L-ISS-10,B-L-ISS-11,B-L-ISS-12},
and that most seesaw extensions of the NMSSM focused on the augmentation of simple Type-I mechanism to study the spectral characters
of gamma-ray from DM annihilations \cite{NMSSM-SS-1,NMSSM-SS-2,NMSSM-SS-3,NMSSM-SS-4,NMSSM-SS-5,NMSSM-SS-6,NMSSM-SS-7,NMSSM-SS-8,NMSSM-SS-9,NMSSM-SS-10}.
These studies usually concentrated on the parameter region which predicts a large DM-nucleon scattering rate and hence has been excluded by current DD experiments. By contrast, only several works have been done to study the theory and phenomenology of the ISS-NMSSM \cite{ISS-NMSSM-1,ISS-NMSSM-2,ISS-NMSSM-3,
ISS-NMSSM-4,ISS-NMSSM-5,ISS-NMSSM-6}. In particular, we note that only the work \cite{ISS-NMSSM-2} adopted same
symmetries as our model, and it studied the effect of $\tilde{\nu}_1$ on the properties of a ${\cal{O}}(10 {\rm GeV})$ CP-odd Higgs boson. This situation necessitates our
study as a helpful attempt to explore the nature of DM. Obviously, our result on DM physics may be distinct from the previous ones since they are based on
different theoretical assumptions and also for different purposes.

This paper is organized as follows. In section 2, we introduce the basics of the ISS-NMSSM, including the annihilation mechanisms of
sneutrino DM and the features of the spin-independent (SI) cross section for DM-nucleon scattering. In Section 3 we scan the parameter
space of the model by considering relevant experimental constraints to get viable parameter points, and analyze numerically the key features of sneutrino DM.
In section 4, we study the constraints of the LHC experiment on our choice of the NMSSM parameters. For this purpose,
we simulate the neutralino/chargino production processes, and point out that current experimental analyses on sparticle search can not exclude the light
higgsino-dominated particles due to their unconventional signatures. Section 5 is devoted to a brief exploration of the phenomenology of the ISS-NMSSM,
and we will show that the phenomenology is quite rich and distinct. Finally, we draw our conclusions in section 6.

\section{NMSSM with Inverse Seesaw Mechanism  }

In this section we first introduce the basics of the ISS-NMSSM,  including its Lagrangian and neutrino physics, then we concentrate on
sneutrino DM case. We analyze the features of sneutrino mass matrix, and present useful formula to calculate the cross sections for DM annihilations
and also that for DM-nucleon scattering.

\subsection{Model Lagrangian}

\begin{table}[t]
\begin{center}
\begin{tabular}{|c|c|c|c|c|c|}
\hline \hline
SF & Spin 0 & Spin \(\frac{1}{2}\) & Generations & \((U(1)\otimes\, \text{SU}(2)\otimes\, \text{SU}(3))\) \\
\hline
\(\hat{q}\) & \(\tilde{q}\) & \(q\) & 3 & \((\frac{1}{6},{\bf 2},{\bf 3}) \) \\
\(\hat{l}\) & \(\tilde{l}\) & \(l\) & 3 & \((-\frac{1}{2},{\bf 2},{\bf 1}) \) \\
\(\hat{H}_d\) & \(H_d\) & \(\tilde{H}_d\) & 1 & \((-\frac{1}{2},{\bf 2},{\bf 1}) \) \\
\(\hat{H}_u\) & \(H_u\) & \(\tilde{H}_u\) & 1 & \((\frac{1}{2},{\bf 2},{\bf 1}) \) \\
\(\hat{d}\) & \(\tilde{d}_R^*\) & \(d_R^*\) & 3 & \((\frac{1}{3},{\bf 1},{\bf \overline{3}}) \) \\
\(\hat{u}\) & \(\tilde{u}_R^*\) & \(u_R^*\) & 3 & \((-\frac{2}{3},{\bf 1},{\bf \overline{3}}) \) \\
\(\hat{e}\) & \(\tilde{e}_R^*\) & \(e_R^*\) & 3 & \((1,{\bf 1},{\bf 1}) \) \\
\(\hat{\nu}\) & \(\tilde{\nu}_R^*\) & \(\nu_R^*\) & 3 & \((0,{\bf 1},{\bf 1}) \) \\
\(\hat{s}\) & \(S\) & \(\tilde{S}\) & 1 & \((0,{\bf 1},{\bf 1}) \) \\
\(\widehat X\) & \(\tilde{x}\) & \(x\) & 3 & \((0,{\bf 1},{\bf 1}) \) \\
\hline \hline
\end{tabular}
\end{center}
\caption{Chiral superfields in the NMSSM with inverse seesaw mechanism.}
\label{table1}
\end{table}

Depending on field assignment and the symmetry adopted in model construction, there are various ways to implement the inverse seesaw mechanism in the NMSSM
\cite{ISS-NMSSM-1,ISS-NMSSM-2,ISS-NMSSM-3,ISS-NMSSM-4,ISS-NMSSM-5,ISS-NMSSM-6}.
Here we consider the minimal framework which extends the NMSSM by only two gauge singlet chiral fields $\nu$ and $X$ for each generation
with lepton numbers $L = -1$ and $L = +1 $ respectively. We assume that the lepton number $L$ and $Z_3$ symmetry are broken slightly, while the $R$-parity
and $(-1)^L$ parity are still good symmetries.  With these assumptions, we write down the theory of the ISS-NMSSM with its field content presented in
Table \ref{table1}, and its superpotential and corresponding soft breaking terms given by \cite{ISS-NMSSM-2}
\begin{eqnarray}
 W &=& \left [ Y_u \,{\hat{q}} \cdot \,\hat{H}_u\,\hat{u} \, + Y_d \,\hat{H}_d\, \cdot \hat{q}\, \hat{d}\, +
 Y_e \,\hat{H}_d \cdot \,\hat{l} \,\hat{e} \,+\lambda\,\hat{s}\,\hat{H}_u \cdot \, \hat{H}_d\,
 +\frac{1}{3} \kappa \,\hat{s}^3 \right ] \nonumber  \\  & & + \left [ \frac{1}{2} \mu_\nu \,\widehat \nu\,\widehat \nu + \frac{1}{2} \mu_X \,\widehat X\,\widehat X\,+
 \lambda_N\,\hat{s}\,\hat{\nu}\,\widehat X\,+Y_\nu \,\hat{l} \cdot \hat{H}_u \,\hat{\nu} \right ],  \label{Superpotential} \\
L^{soft} &=& - \left [ \frac{1}{2}\left(m_1 \lambda_{\tilde{B}}^{2}  + m_2 \lambda_{{\tilde{W}}}^2 + m_3 \lambda_{\tilde{g}}^2 + \mbox{h.c.} \right) +
 m_{H_d}^2 |H_d|^2 +m_{H_u}^2 |H_u|^2 +m_S^2 |S|^2  \right . \nonumber \\
&& \left. + m_{\tilde{q}}^{2} (\tilde{u}^*_{L} \tilde{u}_{L} + \tilde{d}^*_{L} \tilde{d}_{L})
+m_{\tilde{u}}^{2} \tilde{u}^*_{R}  \tilde{u}_{R} + m_{\tilde{d}}^{2} \tilde{d}^*_{R}\tilde{d}_{R} +
 m_{\tilde{l}}^{2} ( \tilde{e}^*_{L} \tilde{e}_{L} + \tilde{\nu}_{L}\tilde{\nu}^*_{L} ) +  m_{\tilde{e}}^{2}
 \tilde{e}^*_{R} \tilde{e}_{R} \right . \nonumber \\
&& \left. +  ( \lambda A_{\lambda} SH_u\cdot H_d +\frac{\kappa}{3} A_{\kappa} S^3 + Y_u A_u \tilde{u}^*_R \tilde{q}\cdot {H}_u +
Y_d A_d\tilde{d}^*_R {H}_d\cdot \tilde{q}+ Y_e A_e\tilde{e}^*_R  H_d \cdot \tilde{l} + h.c. ) \right ]  \nonumber \\
& & - \left [ m_{\nu}^{2} \tilde{\nu}_{R}\tilde{\nu}^*_{R} + m_{x}^{2} \tilde{x}\tilde{x}^* +
(\frac{B_{\mu_\nu}}{2} \tilde{\nu}_R^\ast \tilde{\nu}_R^\ast  + \frac{B_{\mu_X}}{2} \tilde{x} \tilde{x}
+ \lambda_N A_{\lambda_N}S \tilde{\nu}^*_{R} \tilde{x} + Y_\nu A_{\nu} \tilde{\nu}^*_{R} \tilde{l} H_u+ \mbox{h.c.}) \right ]. \nonumber
\end{eqnarray}
In above formulae, the coefficients $\lambda$ and $\kappa$ parameterize the interactions among the Higgs fields, $Y_f$ ($f=u,d,e,\nu$) and $\lambda_N$ are Yukawa couplings
for quarks and leptons, $m_i$ ($i=u, d, \cdots$) denote soft breaking masses, and $A_i$ are soft breaking coefficients for trilinear terms.

About the Lagrangian in Eq.(\ref{Superpotential}), five points should be noted. First, the terms in the first bracket
of the superpotential $W$ correspond to that of the NMSSM with $Z_3$ symmetry \cite{Ellwanger:2009dp}, and those in the second bracket
are for the newly added neutrino superfields. The expression of $L^{soft}$ has same structure. Second, we have neglected flavor indices
in writing down the expressions for the sake of simplicity. So all the parameters except for those in the Higgs and gaugino sectors are
actually $3 \times 3$ (diagonal or non-diagonal) matrices in flavor space. Third, among the parameters in the superpotential
only $\mu_\nu$ and $\mu_X$ are dimensional. These coefficients parameterize the effect of LNV, which may
arise from the integration of heavy particles in an ultraviolet high energy theory with LNV interactions
(see for example \cite{ISS-NMSSM-1,ISS-NMSSM-3} and also discussions in
\cite{ISS-NMSSM-2}), so the magnitude of their elements should be suppressed. Similarly, the coefficients
$B_{\mu_\nu}$ and $B_{\mu_X}$ tend to be small.
Fourth, the fields $H_{u,d}^0$ and $S$ acquire their vevs after electroweak symmetry breaking, i.e.
$ \langle H_u^0 \rangle = v_u/\sqrt{2}$, $ \langle H_d^0 \rangle = v_d/\sqrt{2}$ and $\langle S \rangle = v_s/\sqrt{2}$. These vevs are
related with the soft breaking squared masses $m_{H_u}^2$, $m_{H_d}^2$ and $m_S^2$ by the minimization conditions
of the Higgs potential \cite{Ellwanger:2009dp}, and in practice
one may use $m_Z$, $\tan \beta \equiv v_u/v_d$ and $\mu = \frac{\lambda}{\sqrt{2}} v_s$ instead of the
squared masses as input parameters of the ISS-NMSSM. Finally, we emphasize that the last two terms in the $W$ can induce three/four scalar interactions involving
sneutrinos and Higgs bosons, and their corresponding soft breaking terms induce only three scalar interactions. These interactions,
as we mentioned before, play an important role in DM physics. We also emphasize that the Yukawa coupling $Y_\nu$ can introduce
extra interactions for the superfields $\hat{l}$ and $\hat{H}_u$, and consequently the signature of
left-handed sleptons and higgsinos at the LHC may be altered greatly.

Obviously the Higgs sector of the ISS-NMSSM is same as that of the NMSSM. In this work, we adopt the convention of the NMSSM that
$h_i$ with $i=1,2,3$ ($A_j$ with $j=1,2$) denote mass eigenstates of CP-even Higgs bosons (CP-odd Higgs bosons) with
their mass satisfying $m_{h_1} < m_{h_2} < m_{h_3}$ ($m_{A_1} < m_{A_2}$). Since this sector has been introduced in detail
in \cite{Ellwanger:2009dp}, we in the following only consider the neutrino
and sneutrino sectors. As we will show below, the singlet Higgs fields can play an important role in these sections.

\subsection{Neutrino Sector}

In the ISS-NMSSM,  the neutrino Yukawa interactions take the following form
\begin{eqnarray}
{\cal L}_{\nu} = \nu_R^\ast Y_\nu H_u^0 \nu_L   + \nu_R^\ast \lambda_N S x  + \frac{1}{2} \nu_R^\ast \mu_\nu \nu_R^\ast +
\frac{1}{2} x \mu_X x + {\rm {~h.c.}},
\end{eqnarray}
and they generate the neutrino masses after the electroweak symmetry breaking. In the interaction basis $(\nu_L, \nu_R^\ast, x)$,
the $9\times 9$ neutrino mass matrix reads
\begin{eqnarray}
\label{ISSmatrix}
 M_{\mathrm{ISS}}=\left(\begin{array}{c c c} 0 & M_D^T & 0 \\ M_D & \mu_\nu & M_R \\ 0 & M_R^T & \mu_X \end{array}\right)\,, \label{Neutrino-mass}
\end{eqnarray}
with the $3\times3$ Dirac mass matrices given by $M_D = \frac{v_u}{\sqrt{2}} Y_\nu $ and  $M_R = \frac{v_s}{\sqrt{2}} \lambda_N $.
Since this mass matrix is complex and symmetric, it can be diagonalized by a $9\times 9$ unitary matrix $U_\nu$ according to
\begin{eqnarray}
U^\ast_\nu  M_{\mathrm{ISS}} U^\dag_\nu = {\rm diag}(m_i,m_{H_j}),~ ~(i=1,2,3;~j=1,2,...,6).
\end{eqnarray}
This gives three light neutrino mass eigenstates and six heavy neutrino mass eigenstates, which are related with the
interaction state $\nu$ by $\nu_m = U_\nu \nu$.
Without loss of generality, the matrix $U_\nu^\dag$ can be decomposed into
the blocks
\begin{eqnarray}
\left ( U_{\nu}^\dag \right )_{9\times 9} = \left(\begin{array}{cc}
U_{3\times 3} & X_{3\times 6}\\
Y_{6\times 3} & Z_{6\times 6}
\end{array}\right),  \label{eq:diagfull}
\end{eqnarray}
where the $3 \times 3$ matrix $U$ is responsible for the oscillations of active neutrinos, and the value of its elements can be extracted from relevant
experimental data.

With the definition $\|M\| \equiv \sqrt{{\rm Tr}(M^\dag M)}$ for an arbitrary matrix $M$ and in the limit $\|\mu_X\|, \|\mu_\nu\| \ll \|M_D\| \ll \|M_R\|$,
one can extract the mass matrix of the light active neutrinos from the expression in Eq.(\ref{Neutrino-mass}), which is given by
\begin{eqnarray}
	M_\nu = \left[M_D^T M_R^{T^{-1}}\right]\mu_X \left[(M_R^{-1})M_D\right]+{\cal O}(\mu_{X,\nu}^2) \equiv F \mu_X F^T + {\cal O}(\mu_{X,\nu}^2) \, .
\end{eqnarray}
In above formula, $F = M_D^T M_R^{T^{-1}}$ and the magnitude of its elements is of the order $\|M_D\|/\|M_R\|$.
So in inverse seesaw mechanism, the smallness of the active neutrino masses is not only due to the small elements of the lepton-number violating matrix
$\mu_X$, but also due to the suppression factor $\|M_D\|^2/\|M_R\|^2$. For
$\|\mu_X \| \sim {\cal O}({\rm KeV})$, one can easily get $\|M_R \| \sim {\cal O}$(TeV)
for comparatively large Dirac Yukawa couplings, $ \| Y_\nu \| \sim {\cal O}(0.1)$.
This usually leads to observable lepton flavor violation (LFV) signals as discussed in literatures
\cite{MSSM-ISS-1,MSSM-ISS-3,MSSM-ISS-4,MSSM-ISS-6,MSSM-ISS-14,MSSM-ISS-16,MSSM-ISS-18,Krauss:2013gya}.
Note that although both $\|\mu_X\|$ and $\|\mu_\nu\|$ are naturally small,
$\mu_X$ controls the size of the light neutrino masses, while $\mu_\nu$ is irrelevant. In view of this, for the
sake of simplicity we set the matrix $\mu_\nu$ (and also its soft breaking parameter $B_{\mu_\nu}$) to
be zero and do not discuss its effect any more. Also note that the mass scale of the heavy neutrinos is
determined by the magnitude of $\|M_R \|$.

The symmetric effective light neutrino mass matrix $M_\nu$ can be
diagonalized by the unitary Pontecorvo-Maki-Nakagawa-Sakata (PMNS)
matrix
\begin{eqnarray}
 U_{\rm PMNS}^T M_{\mathrm{\nu}} U_{\rm PMNS} = \mathrm{diag}(m_{\nu_1}\,, m_{\nu_2}\,, m_{\nu_3})\,,
\end{eqnarray}
where $m_{\nu_1}$, $m_{\nu_2}$ and $m_{\nu_3}$ are the masses of the three lightest neutrinos.
Generally speaking, due to the mixings among the states $(\nu_L, \nu_R^\ast, x)$,  the matrix $U$ in
Eq.(\ref{eq:diagfull}) does not coincide with $U_{\rm PMNS}$, instead in the limit $\|\mu_X\|\ll \|M_D\| \ll \|M_R\|$,
they are related by
\begin{eqnarray}
U \simeq \left({\bf 1} - \frac{1}{2}F F^\dag \right)U_{\rm PMNS} \equiv
({\bf 1}-\eta) U_{\rm PMNS}.  \label{non-unitarity}
\end{eqnarray}
In this sense, $\eta=\frac{1}{2}FF^\dag$ is a measure of the non-unitarity of the matrix $U$, which is obtained
from neutrino experiments. On the other hand, since current experiments have tightly limited the violation
of the unitarity \cite{Fernandez-Martinez:2016lgt},  one can assume $U_{\rm PMNS} \simeq U$ and use the data of neutrino
experiments to limit the parameters in $M_{\nu}$. So far two parameterizations schemes are adopted in literature (see for example
\cite{Arganda:2014dta}) in doing this.  One is to express the Yukawa coupling matrix $Y_\nu$ in terms of $U_{\rm PMNS}$ by using a
modified Casas-Ibarra parameterization \cite{Casas:2001sr}, which is given by
\begin{eqnarray}
 m_D = V^\dagger \mathrm{diag}(\sqrt{M_1}\,,\sqrt{M_2}\,,\sqrt{M_3})\; R\; \mathrm{diag}(\sqrt{m_{\nu_1}}\,, \sqrt{m_{\nu_2}}\,, \sqrt{m_{\nu_3}}) U^\dagger_{\rm PMNS}\,.
\end{eqnarray}
Here $V$ is a unitary matrix that diagonalizes $M =M_R \mu_X^{-1} M_R^T$ by
\begin{eqnarray}
 M=V^\dagger \mathrm{diag}(M_1\,, M_2\,, M_3) V^*,
\end{eqnarray}
and $R$ is a complex orthogonal matrix given by
\begin{eqnarray}
R = \left( \begin{array}{ccc} c_{2} c_{3}
& -c_{1} s_{3}-s_1 s_2 c_3& s_{1} s_3- c_1 s_2 c_3\\ c_{2} s_{3} & c_{1} c_{3}-s_{1}s_{2}s_{3} & -s_{1}c_{3}-c_1 s_2 s_3 \\ s_{2}  & s_{1} c_{2} & c_{1}c_{2}\end{array} \right) \,,
\end{eqnarray}
where $c_i\equiv \cos \theta_i$, $s_i\equiv \sin\theta_i$ and
$\theta_1$, $\theta_2$, and $\theta_3$ are arbitrary angles. In this scheme, the neutrino Yukawa coupling $Y_\nu$ is usually flavor non-diagonal.
The other scheme utilizes the fact that once the matrix $Y_\nu$ and $M_R$ are given, $\mu_X$ alone can be responsible for neutrino experimental data.
In this case, $\mu_X$ is given by \cite{Arganda:2014dta,Baglio:2016bop}
\begin{eqnarray}
\mu_X=M_R^T ~m_D^{T^{-1}}~ U_{\rm PMNS}^* \, \mathrm{diag}(m_{\nu_1}\,, m_{\nu_2}\,, m_{\nu_3})\,  U_{\rm PMNS}^\dagger~ {m_D}^{-1} M_R. \label{mu-X}
\end{eqnarray}
Note that for this scheme, one may set $Y_\nu$ and $\lambda_N$ to be flavor diagonal, and
this choice can simplify greatly our study on the properties of sneutrino DM (see following
discussion about sneutrino mass matrix).

\subsection{Sneutrino Dark Matter}

In the ISS-NMSSM, the lightest sneutrino  $\tilde{\nu}_1$ may be a better DM candidate than the customary lightest neutralino after considering
the negative result in recent DM DD experiments,  which is the main standpoint of this work. In the following, we will present in detail the
features of $\tilde{\nu}_1$, including its mass, its annihilation channels as well as its scattering with nucleon.

\subsubsection{Sneutrino mass matrices}

After decomposing sneutrino fields into CP-even and CP-odd parts
\begin{eqnarray}
\tilde{\nu}_{L,i} =  \, \frac{1}{\sqrt{2}} \left ( \phi_i  + i \sigma_i \right ),~~~~
\tilde{\nu}_{R,i} = \frac{1}{\sqrt{2}} \left (\phi_{3+i}  + i \sigma_{3 + i} \right ), ~~~~ \tilde{x}_i = \frac{1}{\sqrt{2}}
\left ( \phi_{6 + i}  + i \sigma_{6+i} \right ),
\end{eqnarray}
with $i=1,2,3$ representing flavor index, one can write down the mass matrix for the CP-odd sneutrinos in the basis
$\sigma_j$ ($j=1,\cdots 9$) as follows
\begin{eqnarray}
m^2_{\tilde{\nu}_I} = \left(
\begin{array}{ccc}
m_{11} \quad &m_{12} \quad & m_{13} \\
m_{12}^T \quad  &m_{22} \quad &m_{23}\\
m_{13}^T \quad &m_{23}^T \quad &m_{33} \end{array}
\right),
\label{cp-odd}
 \end{eqnarray}
where
\begin{eqnarray}
m_{11} &=& \frac{1}{4} \left [ 2 v_{u}^{2} {\Re\Big({Y_{\nu}^{T}  Y_\nu^*}\Big)}  + 4 {\Re\Big(m_l^2\Big)} \right ]  + \frac{1}{8} \Big(g_{1}^{2} +
g_{2}^{2}\Big) \Big(- v_{u}^{2}  + v_{d}^{2}\Big) {\bf 1},  \nonumber \\
m_{12} &=& -\frac{1}{2} v_d v_s {\Re\Big(\lambda Y_\nu^* \Big)}  + \frac{1}{\sqrt{2}} v_u {\Re\Big(Y_\nu A_\nu \Big)}, \nonumber  \\
m_{13} &=& \frac{1}{2} v_s v_u {\Re\Big({Y_{\nu}^{T}  \lambda_N^*}\Big)}, \nonumber \\
m_{22} &=& \frac{1}{4} \left [ 2 v_{s}^{2} {\Re\Big({\lambda_N  \lambda_{N}^{\dagger}}\Big)}  + 2 v_{u}^{2} {\Re\Big({Y_\nu  Y_{\nu}^{\dagger}}\Big)}
+ 4 {\Re\Big(m_{\nu}^2\Big)} \right ],  \nonumber \\
m_{23} &=& \frac{1}{8} \left \{ -2 v_d v_u \lambda^* \lambda_{N}^{T}  + 2 \left [ \Big(- v_d v_u \lambda  + v_{s}^{2} \kappa \Big)\lambda_{N}^{\dagger}  + v_{s}^{2} \kappa^* \lambda_{N}^{T} \right ] \right . \nonumber \\
&& \quad \quad \quad \quad \quad \quad \quad \quad \quad \quad \quad \quad \left .  + \sqrt{2} v_s \left [ -4 {\Re\Big({\mu_X  \lambda_{N}^{\dagger}}\Big)}
+ 4 {\Re\Big(A_{\lambda_N}^{T} \lambda_N^T \Big)} \right ] \right \}, \nonumber \\
m_{33} &=& \frac{1}{8} \Big(4 v_{s}^{2} {\Re\Big({\lambda_{N}^{T}  \lambda_N^*}\Big)}  -8 {\Re\Big(B_{\mu_X}\Big)}  + 8 {\Re\Big({\mu_X  \mu_X^*}\Big)}  + 8 {\Re\Big(m_x^2\Big)} \Big),
\end{eqnarray}
and all the $m_{ij}$  are $3 \times 3 $ matrices in flavor space. From the expression of $m_{\tilde{\nu}_I}^2$, one can get following conclusions
\begin{itemize}
\item In the case of no flavor mixing in the matrix $m_{ij}$, which can be obtained by neglecting the small flavor non-diagonal matrix $\mu_X$ presented in Eq.(\ref{mu-X}) (and also the coefficient of the bilinear term $B_{\mu_X}$) and is the situation considered in this work,
one can rearrange the basis $\sigma_j$ by the order $(\sigma_1,\sigma_4,\sigma_7,\sigma_2,\sigma_5,\sigma_8,\sigma_3,\sigma_6,\sigma_9)$ so that
$m^2_{\tilde{\nu}_I}$ is flavor diagonal. In this case,
there are only the mixings between $(\tilde{\nu}_L, \tilde{\nu}_R, \tilde{x})$ for same generation sneutrinos. If the lightest
sneutrino comes from a certain generation, e.g. the third generation, and at same time it is significantly lighter than the other
generation sneutrinos,  one only needs to consider the mass matrix for this generation
sneutrinos in discussing the properties of the DM \footnote{We checked that for the case of mass-degenerate sneutrino DM with different flavors, the relic density will be increased in comparison with the non-degenerate case. This effect, however, can be compensated for by the reduced couplings in DM annihilation. }. This will greatly simply our analysis.
In the following, we only consider one generation of sneutrinos in studying the property of $\tilde{\nu}_1$.

\item Among the parameters in sneutrino sector, $Y_\nu$, $A_\nu$, $\lambda_N$ and $A_{\lambda_N}$ affect not only the interactions of
the sneutrinos, but also the mass spectrum of the sneutrinos. By contrast, the soft breaking masses $m_{\nu}^2$ and $m_{x}^2$
and the small bilinear coefficient $B_{\mu_\nu}$ only affect the spectrum. Considering that the former four parameters are tightly limited
by various experiments (see below), one can conclude that the spectrum is mainly determined by the soft breaking masses for heavy
sneutrino case; on the other hand, since $v_s$ is usually much larger than $v_u$, the spectrum is more sensitive to $\lambda_N$ and $A_{\lambda_N}$
than to the other parameters for the case of light sneutrinos, $m_{\tilde{\nu}_i} \sim v_u$.
\item The mixing of $\tilde{\nu}_L$ with the other fields is determined by the parameters $Y_\nu$ and $A_\nu$. ¡¡
In the limit $Y_{\nu} = 0$, $m_{12}$ and $m_{13}$ vanish, and consequently $\tilde{\nu}_L$ does not mix
with $\tilde{\nu}_R$ and $\tilde{x}$ any more. Furthermore, if the first term in $m_{22}$ is far dominant over the rest terms in $m_{22}$ and so is $m_{33}$, $m_{22} \simeq m_{33}$ and this results in a maximal mixing between $\tilde{\nu}_R$ and $\tilde{x}$. In this case, $\tilde{\nu}_1$ is approximated by
    $\tilde{\nu}_1 \simeq 1/\sqrt{2} [ Im(\tilde{\nu}_R) - Sgn(m_{23}) Im(\tilde{x}) ] $.
    This situation is frequently encountered in our results.
\end{itemize}

In a similar way one may discuss the mass spectrum of the CP-even sneutrinos. We find that their mass matrix $m^2_{\tilde{\nu}_R}$ is related with $m^2_{\tilde{\nu}_I}$ by $m^2_{\tilde{\nu}_R} = m^2_{\tilde{\nu}_I} |_{\mu_X \to - \mu_X, B_{\mu_X} \to - B_{\mu_X}}$. Since the Majorana mass
$\mu_X$ and the bilinear coefficient $B_{\mu_X}$ reflect the effect of LNV, they should be suppressed greatly. In the limit $\mu_X = 0$ and $B_{\mu_X} = 0$,
any CP-even sneutrino particle
must be accompanied with a mass-degenerate CP-odd sneutrino. In this case, one may say that the sneutrino as an mass eigenstate corresponds to a complex field, and it has
its anti-particle \cite{MSSM-ISS-10}. If alternatively $B_{\mu_X}$ takes a moderately small value and consequently the mass splitting between the CP-even sneutrino particle and its
corresponding CP-odd particle is at ${\rm eV}$ order, one may call such a sneutrino pseudo-complex particle.
This case has interesting implication in DM physics \cite{MSSM-ISS-15,ISS-NMSSM-3}.

In this work, we only consider the case that $B_{\mu_X}$ is moderately large, $B_{\mu_X} = 20 {\rm GeV^2}$, so that the CP-odd state is lighter than its corresponding CP-even
state by $\sim 0.1 \rm GeV$, and sneutrinos as mass eigenstates have definite CP number. We note that the lightest CP-even sneutrino $\tilde{\nu}_1^R$
can decay into $\tilde{\nu}_1 \gamma $  with a width around the order of $10^{-8} {\rm GeV}$, and it usually coannihilated  in early Universe with $\tilde{\nu}_1$
to get the right DM relic density.   We numerically checked by the code micrOMEGAs \cite{micrOMEGAs-1,micrOMEGAs-2,micrOMEGAs-3} that the observables such as the relic density and DM-nucleon scattering rate discussed in this work are insensitive to the choice of $B_{\mu_X}$.

\subsubsection{Relic density of sneutrino DM}

In the cosmological standard model, the abundance of a thermal DM $Y(T)$ is defined as the number density divided by entropy density $s(T)$,
and its Boltzmann equation is ~\cite{Belanger:2001fz}
\begin{eqnarray}
\frac{dY}{dT}=\sqrt{\frac{\pi g_{\ast}(T)}{45}}M_{p} -\left\langle \sigma v\right\rangle(Y^{2}-Y^{2}_{eq})\,,
\label{relicabundance}
\end{eqnarray}
where $g_{\ast}$ is an effective number of degrees of freedom (dof) derived from
thermodynamics describing state of the Universe,  $M_{p}$ is Plank mass,  $Y_{eq}$ is thermal equilibrium abundance,
and $\left\langle \sigma v\right\rangle$ is the relativistic thermally averaged annihilation cross section with $v$  denoting
the relative velocity between the annihilating particles. With the aid of present day abundance $Y(T_0)$,
the DM density today can be written as ~\cite{Belanger:2001fz}
\begin{eqnarray}
\Omega h^{2}= \frac{8 \pi}{3} m_{DM} \frac{s(T_0) Y(T_0)}{M_p^2 (100 ({\rm km}/{\rm s}/{\rm Mpc}))^2} = 2.742\times 10^{8} \times \frac{m_{DM}}{\mbox{GeV}} \times Y(T_{0})\,,
\label{solution}
\end{eqnarray}
where $s(T_0)$ is the entropy density at present time and $h$ is the normalized Hubble constant. These formulae indicate that
in order to get the right relic density, one has to solve the evolution equation of $Y(T)$, which is usually a complicated
work and has to be done numerically.

\begin{figure}[t]
\centering
\includegraphics[height=3cm, width=12cm]{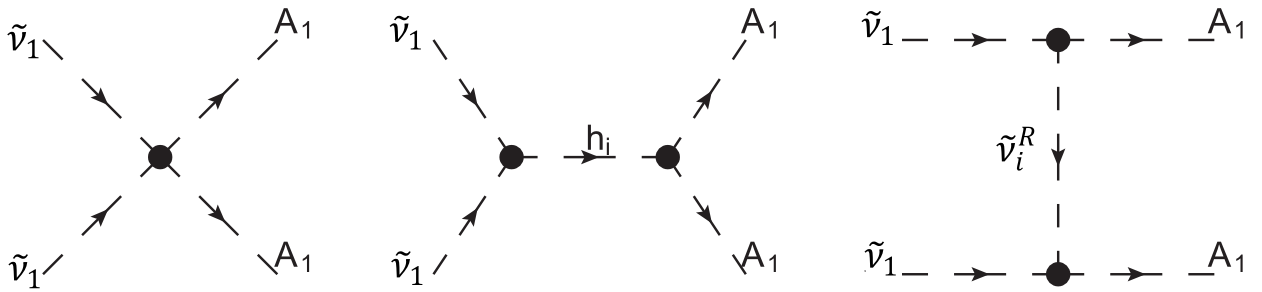}
\caption{Feynman diagrams for the annihilation $\tilde{\nu}_1 \tilde{\nu}_1 \to A_1 A_1$. Note that there exist both $t$ and $u$ channel contribution to the annihilation in the right diagram. \label{fig1}}
\end{figure}

As far as the ISS-NMSSM is concerned, its influence on the relic density of $\tilde{\nu}_1$ enters
through the cross section $\left\langle \sigma v\right\rangle$, which includes all annihilation and coannihilation channels predicted by the model,
and is given by ~\cite{Belanger:2001fz}
\begin{eqnarray}
\left\langle \sigma v\right\rangle=\frac{\displaystyle\sum_{i,j}g_{i}g_{j}\displaystyle\int_{ (m_{i}+m_{j})^{2} }ds\sqrt{s}K_{1}(\frac{\sqrt{s}}{T})p_{ij}^{2}\displaystyle\sum_{k,l}\sigma_{ij;kl}(s)}{2T(\displaystyle\sum_{i}g_{i}m^{2}_{i}K_{2}(m_{i}/T) )^{2}}\,,
\label{evolutionabundance}
\end{eqnarray}
where $g_{i}$ is the number of dof, $\sigma_{ij;kl}$ is the cross section for the annihilation of a pair of supersymmetric particles
with masses $m_{i}$, $m_{j}$ into SM particles $k$ and $l$, $p_{ij}$ is the momentum of incoming particles in their center
of mass frame with total energy $\sqrt{s}$, and $K_1$ and $K_2$ are modified Bessel functions. In practice, the potentially important contributions
to $\left \langle \sigma v\right\rangle$ include following annihilation channels
\begin{itemize}
\item[(1)] $\tilde{\nu}_1 \tilde{\nu}_1 \to s s$ with $s$ denoting either a CP-even or CP-odd singlet dominant Higgs boson.
This annihilation proceeds via a four-point scalar coupling, s-channel mediation of a Higgs boson and $t$/$u$- exchange of
a sneutrino, which are depicted in Fig.\ref{fig1} for the case that $s$ is
the lightest CP-odd Higgs $A_1$.
\item[(2)] $\tilde{\nu}_1 \tilde{\nu}_1 \to \eta \eta^\ast$ with $\eta$ denoting a SM particle or any of the heavy neutrinos.
This annihilation is mediated by any of the CP-even Higgs bosons, and since the involved interactions are usually weak
in getting the right relic density, one of the bosons must be at resonance.
\item[(3)] $\tilde{\nu}_1^R \tilde{\nu}_1^R \to s s, \eta \eta^\ast$ and $\tilde{\nu}_1 \tilde{\nu}^R_1 \to A^{(*)} \to \eta \eta^{\ast}$
which are similar to the channels (1) and (2). Note that $\tilde{\nu}_1^R$ plays an important role in determining the relic
density since $\tilde{\nu}_1^R$ is always nearly degenerate with $\tilde{\nu}_1$ in mass.

\item[(4)] $\tilde{\chi}_i^0 \tilde{\chi}_j^0, \tilde{\chi}_i^0 \tilde{\chi}_1^\pm, \tilde{\chi}_1^\pm  \tilde{\chi}_1^\mp \to \eta_i \eta^{\ast}_j$ with $\tilde{\chi}$ denoting a higgsino-like or wino-like electroweakino. These annihilations are called coannihilation in literature \cite{Coannihilation,Griest:1990kh}, and to make the effect significant, the
    mass splitting between $\tilde{\chi}$ and $\tilde{\nu}_1$ should be less than about $10\%$.
\end{itemize}

In the following, we consider for illustration purpose the cross section of the annihilation channel shown in Fig.\ref{fig1} with collision energy $\sqrt{s}$, which is given by
\begin{eqnarray}
\sigma v |_{\sqrt{s}} &=&  \frac{\sqrt{1- m_{A_1}^2/s}}{16 \pi s} \left \{ \left | C_{\tilde{\nu}_1 \tilde{\nu}_1 A_1 A_1} -
\sum_i \frac{C_{\tilde{\nu}_1\tilde{\nu}_1 h_i } C_{h_i A_1 A_1}}{s -m_{h_i}^2} \right |^2  \right . \nonumber \\
&& \left .
+ 2 C_{\tilde{\nu}_1 \tilde{\nu}_1 A_1 A_1} \sum_{i} \frac{C_{\tilde{\nu}_1 \tilde{\nu}^R_i A_1}^2}{c} \ln \frac{a_i + c}{a_i - c}- 2 \sum_{i,j}
\frac{C_{\tilde{\nu}_1\tilde{\nu}_1 h_i } C_{h_i A_1 A_1}}{s -m_{h_i}^2}
\frac{C_{\tilde{\nu}_1 \tilde{\nu}^R_j A_1}^2}{c} \ln \frac{a_j + c}{a_j - c} \right .
\nonumber \\ && \left .+ 2 \sum_{i,j}  \frac{C_{\tilde{\nu}_1 \tilde{\nu}^R_i A_1}^2
C_{\tilde{\nu}_1 \tilde{\nu}^R_j A_1}^2}{c (a_i - a_j)} \ln \frac{(a_i -c ) (a_j+c)}{(a_i + c) (a_j -c)}   \right \} \nonumber \\
& \simeq & a + b v^2, \label{Expression-1} \\
a&=&  \frac{\sqrt{1- m_{A_1}^2/m_{\tilde{\nu}_1}^2}}{64 \pi m_{\tilde{\nu}_1}^2} \left | C_{\tilde{\nu}_1 \tilde{\nu}_1 A_1 A_1} -
\sum_i \frac{C_{\tilde{\nu}_1\tilde{\nu}_1 h_i } C_{h_i A_1 A_1}}{ 4 m_{\tilde{\nu}_1}^2  -m_{h_i}^2} +
\sum_j \frac{2 C_{\tilde{\nu}_1 \tilde{\nu}^R_j A_1}^2}{m_{\tilde{\nu}_1}^2 + m_{\tilde{\nu}_j^R}^2 - m_{A_1}^2} \right |^2,   \\
b&=& \left ( -\frac{1}{4} + \frac{m_{A_1}^2}{8 (m_{\tilde{\nu}_1}^2 - m_{A_1}^2 )} \right ) \times a -
\frac{\sqrt{1- m_{A_1}^2/m_{\tilde{\nu}_1}^2}}{64 \pi m_{\tilde{\nu}_1}^2} \times \nonumber \\
&& \left \{ \sum_{i,j} \frac{C_{\tilde{\nu}_1 \tilde{\nu}_1 h_i} C_{h_i A_1 A_1} C_{\tilde{\nu}_1 \tilde{\nu}_1 h_j} C_{h_j A_1 A_1}}{(4 m_{\tilde{\nu}_1}^2 - m_{h_i}^2) (4 m_{\tilde{\nu}_1}^2 - m_{h_j}^2)} \left ( \frac{m_{\tilde{\nu}_1}^2}{4 m_{\tilde{\nu}_1}^2 - m_{h_i}^2} +   i\leftrightarrow j \right )\right . \nonumber \\
&& \left . - \sum_{i,j} \frac{2 C_{\tilde{\nu}_1 \tilde{\nu}_1 h_i} C_{h_i A_1 A_1} C_{\tilde{\nu}_1 \tilde{\nu}_j^R A_1}^2} {(4 m_{\tilde{\nu}_1}^2 - m_{h_i}^2) ( m_{\tilde{\nu}_1}^2 +  m_{\tilde{\nu}_j^R}^2- m_{A_1}^2)} \left ( \frac{m_{\tilde{\nu}_1}^2}{4 m_{\tilde{\nu}_1}^2 - m_{h_i}^2} +  \frac{2m_{\tilde{\nu}_1}^2}{ m_{\tilde{\nu}_1}^2 +  m_{\tilde{\nu}_j^R}^2 - m_{A_1}^2} \right )\right . \nonumber \\
&& \left . + \sum_{i,j} \frac{2 C_{\tilde{\nu}_1 \tilde{\nu}_i^R A_1}^2 C_{\tilde{\nu}_1 \tilde{\nu}_j^R A_1}^2}{ ( m_{\tilde{\nu}_1}^2 +  m_{\tilde{\nu}_i^R}^2- m_{A_1}^2) ( m_{\tilde{\nu}_1}^2 +  m_{\tilde{\nu}_j^R}^2- m_{A_1}^2)} \left ( \frac{m_{\tilde{\nu}_1}^2}{( m_{\tilde{\nu}_1}^2 +  m_{\tilde{\nu}_i^R}^2- m_{A_1}^2)} +  i\leftrightarrow j \right )\right . \nonumber \\
&& \left . - 2 C_{\tilde{\nu}_1 \tilde{\nu}_1 A_1 A_1}  \left ( \sum_{i} \frac{ C_{\tilde{\nu}_1 \tilde{\nu}_1 h_i} C_{h_i A_1 A_1} m_{\tilde{\nu}_1}^2 } {(4 m_{\tilde{\nu}_1}^2 - m_{h_i}^2)^2 } -  \sum_j \frac{ C_{\tilde{\nu}_1 \tilde{\nu}_j^R A_1}^2  m_{\tilde{\nu}_1}^2 } {( m_{\tilde{\nu}_1}^2 +  m_{\tilde{\nu}_j^R}^2- m_{A_1}^2)^2 } \right ) \right \}.
\end{eqnarray}
In above formulae, $a_i = \frac{s}{2} + m_{\tilde{\nu}^R_i}^2 - m_{\tilde{\nu}_1}^2 - m_{A_1}^2 $,
$c = \frac{1}{2} \sqrt{(s- 4 m_{\tilde{\nu}_1}^2) (s - 4 m_{A_1}^2)}$, $C_{\tilde{\nu}_1 \tilde{\nu}_1 A_1 A_1}$
denotes the coupling of two $\tilde{\nu}_1$s with two $A_1$s which is mainly determined by the parameter $\lambda_N$,
and the other coefficients $C_{XYZ}$ are the triple scalar couplings involving the particles $X$, $Y$ and $Z$.
In getting the approximation in Eq.(\ref{Expression-1}),
we use the relation $s = 16 m_{\tilde{\nu}_1}^2/(4 - v^2)$ with $v$ denoting the relative velocity of the two $\tilde{\nu}_1$s, and assume following conditions:
(1) $v \sim {\cal{O}}(0.1)$, which means that the collision is nonrelativistic;  (2) $\tilde{\nu}_1$ is much heavier than $m_{A_1}$; (3)
$m_{\tilde{\nu}_1}/|2 m_{\tilde{\nu}_1} - m_{h_i}|$ is at most a ${\cal{O}}(1)$ quantity, which excludes the possibility that the mediating
Higgs boson is resonant. With these conditions, the coefficient $b$ is usually smaller than the coefficient $a$.

The thermal averaged cross section of the annihilation at freeze-out temperature $T_f$  and that at present time are then given
by \cite{Griest:1990kh}
\begin{eqnarray}
\left\langle \sigma v\right\rangle_{T_f} \simeq a + 6 b\frac{T_f}{m_{\tilde{\nu}_1}} \simeq a + \frac{6}{25} b \simeq a, \quad \quad \left\langle \sigma v\right\rangle_{0} \simeq a.
\end{eqnarray}
This implies that, if the annihilation $\tilde{\nu}_1 \tilde{\nu}_1 \to A_1 A_1 $ is fully responsible for current relic density
so that $\left\langle \sigma v\right\rangle_{T_f} \sim 3\times 10^{-26} {\rm cm^3 s^{-1}}$,
$\left\langle \sigma v\right\rangle_{0} \simeq \left\langle \sigma v\right\rangle_{T_f} \sim 10^{-26} {\rm cm^3 s^{-1}}$.
Obviously, such a large $\left\langle \sigma v\right\rangle_{0}$ is tightly limited by the Fermi-LAT search for DM annihilation
from dwarf spheroidal galaxy (dSph). In order to avoid the constraint,
one may consider following cases as pointed out by the classical paper \cite{Griest:1990kh}
\begin{itemize}
\item Coannihilation, or more general mixed annihilations. In this case, the annihilation $\tilde{\nu}_1 \tilde{\nu}_1 \to A_1 A_1$ plays a minor role in contributing
to the relic density, and consequently $\left\langle \sigma v\right\rangle_{0}$ can be lowered significantly.
\item $\tilde{\nu}_1$ is slightly lighter than $A_1$, which is called forbidden annihilation in \cite{Griest:1990kh}. In this case,
since the freeze-out occurs at a temperature $T_f \simeq m_{\tilde{\nu}_1}/25$, and also since the velocity of $\tilde{\nu}_1$
is Boltzmann distributed with the temperature, the annihilation
may proceed in early Universe, but can not occur at present time. So $\left\langle \sigma v\right\rangle_{0}$ is suppressed greatly.
\item Resonant annihilation mediated by $h_i$ as the main contribution to the relic density. In this case,  $\left\langle \sigma v\right\rangle_{0}$ can be significantly lower than
$\left\langle \sigma v\right\rangle_{T_f}$ if $2 m_{\tilde{\nu}_1} < m_{h_i}$ \cite{Griest:1990kh,Cao:2014efa}.
\end{itemize}
As we will show below, these cases are frequently  encountered in our scan over the parameter space of the ISS-NMSSM to escape the constraints from the dSph.

Throughout this work, we use the package micrOMEGAs \cite{micrOMEGAs-1,micrOMEGAs-2,micrOMEGAs-3} to evaluate observables in DM physics, including the relic density,
photon spectrum from DM annihilation in the dSph which is used for DM indirect detections, and also the cross sections for DM-nucleon scattering.
The package solves the equation for the abundance
in Eq.(\ref{relicabundance}) numerically without any approximation. In addition, it also estimates
the relative contribution of each individual annihilation or coannihilation channel to the
relic density at the freeze-out temperature.

\subsubsection{Direct detection}

Since the DM $\tilde{\nu}_1$ in the ISS-NMSSM is a scalar with certain lepton and CP numbers, its interaction with nucleon $N$ ($N=p,n$) is mediated only by CP-even Higgs bosons
to result in the effective operator ${\cal{L}}_{\tilde{\nu}_1 N} = f_N \tilde{\nu}_1 \tilde{\nu}_1 \bar{\psi}_N \psi_N$, where the coefficient $f_N$ is  \cite{Han:1997wn}
\begin{eqnarray*}
f_N &=&  m_N \sum_{i=1}^3 \frac{C_{\tilde{\nu}_1 \tilde{\nu}_1 h_i} C_{h_i N N}}{m_{h_i}^2} =  m_N \sum_{i=1}^3 \frac{C_{\tilde{\nu}_1 \tilde{\nu}_1 h_i}}{m_{h_i}^2} \frac{(-g)}{2 m_W} \left ( \frac{S_{i2}}{\sin\beta} F^{(N)}_u +  \frac{S_{i1}}{\cos \beta} F^{(N)}_d \right ).
\end{eqnarray*}
In this formula, $C_{h_i N N}$ is the Yukawa coupling of the Higgs boson $h_i$ with nucleon $N$, $S_{ij}$ is the $(i,j)$ element of the matrix $S$ which is used to diagonalize the CP-even Higgs mass matrix in the basis $(H_d, H_u, s)$,   $F^{(N)}_u=f^{(N)}_u+\frac{4}{27}f^{(N)}_G$ and $F^{(N)}_d=f^{(N)}_d+f^{(N)}_s+\frac{2}{27}f^{(N)}_G$ are nucleon form factors
with $f^{(N)}_q=m_N^{-1}\left<N|m_qq\bar{q}|N\right>$
(for $q=u,d,s$) and $f^{(N)}_G=1-\sum_{q=u,d,s}f^{(N)}_q$. This operator indicates that the spin-dependent cross section for $\tilde{\nu}_1$
scattering with proton vanishes, whereas the SI cross section is given by  \cite{Han:1997wn}
\begin{eqnarray}
\sigma^{\rm SI}_{\tilde{\nu}_1-p} &=& \frac{\mu^2_{\rm red}}{4 \pi m_{\tilde{\nu}_1}^2} f_p^2 = \frac{4 F^{(p)2}_u \mu^2_{\rm red} m_p^2}{\pi}
\left \{ \sum_i ({a_{u}}_i  + {a_{d}}_i F^{(p)}_d/F^{(p)}_u  ) \right \}^2,
 \label{SI-expression}
\end{eqnarray}
where $\mu_{\rm red}= m_p/( 1+ m_p^2/m_{\tilde{\nu}_1}^2) $ is the reduced mass of proton with $m_{\tilde{\nu}_1}$, and the quantities
${a_{u}}_i$ and ${a_{d}}_i$ are defined by
\begin{eqnarray}
{a_{u}}_i = -\frac{g}{8 m_W} \frac{C_{\tilde{\nu}_1 \tilde{\nu}_1 h_i}}{m_{h_i}^2 m_{\tilde{\nu}_1}} \frac{S_{i2}}{\sin\beta}, \quad \quad
{a_{d}}_i = -\frac{g}{8 m_W} \frac{C_{\tilde{\nu}_1 \tilde{\nu}_1 h_i}}{m_{h_i}^2 m_{\tilde{\nu}_1}} \frac{S_{i1}}{\cos \beta},
\end{eqnarray}
to facilitate our analysis. By contrast, we note that ${a_u}_i$ and ${a_d}_i$ in the MSSM with the lightest neutralino $\tilde{\chi}_1^0$ acting as DM
candidate take following form \cite{NMSSM-SS-10}
\begin{eqnarray}
{a_u}_i = - \frac{g}{4 m_W} \frac{C_{\tilde{\chi}_1^0 \tilde{\chi}_1^0 h_i}}{m_{h_i}^2} \frac{S_{i2}}{\sin\beta}, \quad \quad
{a_d}_i = - \frac{g}{4 m_W} \frac{C_{\tilde{\chi}_1^0 \tilde{\chi}_1^0 h_i}}{m_{h_i}^2} \frac{S_{i1}}{\cos \beta},
\end{eqnarray}
where $C_{\tilde{\chi}_1^0 \tilde{\chi}_1^0 h_i} = g_1 N_{11} ( S_{i1} N_{13} - S_{i2} N_{14})$ is the coupling coefficient of
the $\tilde{\chi}_1^0 \tilde{\chi}_1^0 h_i$ interaction with $N$ denoting the rotation matrix to diagonalize neutralino mass matrix in the MSSM.
This implies that
\begin{eqnarray}
\frac{{a_q}_i^{\rm ISS-NMSSM}}{{a_q}_i^{\rm MSSM}} = \frac{1}{2} \frac{C_{\tilde{\nu}_1 \tilde{\nu}_1 h_i}}{C_{\tilde{\chi}_1^0 \tilde{\chi}_1^0 h_i} m_{\tilde{\nu}_1}}.
\label{Ratio}
\end{eqnarray}
We will return to this issue later.

In our numerical calculation of  $\sigma^{\rm SI}_{\tilde{\nu}_1-p}$, we use the default setting of the package micrOMEGAs \cite{micrOMEGAs-1,micrOMEGAs-2,micrOMEGAs-3}
for the nucleon form factors, $\sigma_{\pi N} = 34 {\rm MeV}$ and $\sigma_0 = 42 {\rm MeV}$, and
obtain $F_u^{(p)} \simeq 0.15$ and  $F_d^{(p)} \simeq 0.14$ \footnote{We remind that different choices of $\sigma_{\pi N}$ and $\sigma_0$ can induce an
uncertainty of ${\cal{O}} (10\%)$ on $F_u^{(p)}$ and $F_d^{(p)}$.  For example, if we take  $\sigma_{\pi N} = 59 {\rm MeV}$ and $\sigma_0 = 57 {\rm MeV}$,
which are determined from \cite{Alarcon:2011zs} and \cite{Alarcon:2012nr} respectively, we obtain $F_u^{(p)} \simeq 0.16$ and  $F_d^{(p)} \simeq 0.13$.}. In this case,
Eq.(\ref{SI-expression}) can be approximated by
\begin{eqnarray}
\sigma^{\rm SI}_{\tilde{\nu}_1-p} & \simeq &  \frac{4 F^{(p)2}_u \mu^2_{\rm red} m_p^2  }{\pi}
\left \{ \frac{g}{8 m_W} \sum_i \left [ \frac{C_{\tilde{\nu}_1 \tilde{\nu}_1 h_i}}{m_{h_i}^2 m_{\tilde{\nu}_1}} (\frac{S_{i2}}{\sin\beta} + \frac{S_{i1}}{\cos \beta} ) \right ] \right \}^2
 \label{SI-simplify}  \\
&=& \frac{g^2 F^{(p)2}_u  \mu^2_{\rm red} m_p^2}{16 \pi m_W^2}
\left \{ \sum_i \left [ \frac{(S_{i1} C_{\tilde{\nu}_1 \tilde{\nu}_1 H_d} + S_{i2} C_{\tilde{\nu}_1 \tilde{\nu}_1 H_u} + S_{i3} C_{\tilde{\nu}_1 \tilde{\nu}_1 s} ) }{m_{h_i}^2 m_{\tilde{\nu}_1}} (\frac{S_{i2}}{\sin\beta} + \frac{S_{i1}}{\cos \beta} ) \right ] \right \}^2,  \nonumber
\end{eqnarray}
where $C_{\tilde{\nu}_1 \tilde{\nu}_1 S}$ ($S=H_d, H_u, s$) denotes the coupling of $\tilde{\nu}_1$ with the scalar field $S$, and  for one generation sneutrino case it is given by
\begin{eqnarray}
C_{\tilde{\nu}_1 \tilde{\nu}_1 H_d} &=& \lambda Y_\nu v_s Z_{11} Z_{12} + \lambda \lambda_N v_u Z_{12} Z_{13} - \frac{1}{4} (g_1^2 + g_2^2) v_d Z_{11} Z_{11}, \nonumber \\
C_{\tilde{\nu}_1 \tilde{\nu}_1 H_u} &=&  \lambda \lambda_N v_d Z_{12} Z_{13} - \sqrt{2} T_\nu Z_{11} Z_{12} - Y_\nu^2 v_u  Z_{11} Z_{11} - \lambda_N Y_\nu v_s Z_{11} Z_{13} \nonumber \\
&& \quad \quad - Y_\nu^2 v_u Z_{12} Z_{12} + \frac{1}{4} (g_1^2 + g_2^2) v_u Z_{11} Z_{11},  \nonumber \\
C_{\tilde{\nu}_1 \tilde{\nu}_1 s} &=& \lambda Y_\nu v_d Z_{11} Z_{12} - 2 \kappa \lambda_N v_s  Z_{12} Z_{13} - \sqrt{2} T_{\lambda_N} Z_{12} Z_{13}  + \sqrt{2} \lambda_N \mu_X Z_{12} Z_{13} \nonumber \\ && \quad \quad -  \lambda_N Y_\nu v_u Z_{11} Z_{13} - \lambda_N^2 v_s (Z_{12} Z_{12} + Z_{13} Z_{13} ),
\end{eqnarray}
with $T_\nu \equiv Y_\nu A_\nu$, $T_{\lambda_N} \equiv \lambda_N A_{\lambda_N}$, $Z$ denoting the rotation matrix to diagonalize the CP-odd sneutrino mass matrix and consequently
$\tilde{\nu}_i = Z_{i1} \tilde{\nu}_L + Z_{i2} \tilde{\nu}_R + Z_{i3} \tilde{x}$.

In the following, we analyze the features of $\sigma^{\rm SI}_{\tilde{\nu}_1-p}$. From Eq.(\ref{SI-simplify}), we learn that the dependence of $\sigma^{\rm SI}_{\tilde{\nu}_1-p}$
on the parameters of the ISS-NMSSM comes from the expression in the bracket, which is quite complicated. To simplify the analysis, we assume that
the left-handed sneutrino component in $\tilde{\nu}_1$ is suppressed greatly, e.g. $|Z_{11}| \lesssim 0.05$, and that
\begin{eqnarray}
\tan \beta \gg 1, \quad  Y_\nu, \kappa, \lambda, \lambda_N \sim {\cal{O}}(0.1), \quad  v_s, A_\nu, A_{\lambda_N} \sim 1 {\rm TeV}. \nonumber
\end{eqnarray}
Then the couplings $C_{\tilde{\nu}_1 \tilde{\nu}_1 S}$ can be approximated by
\begin{eqnarray}
C_{\tilde{\nu}_1 \tilde{\nu}_1 H_d} &\simeq & \lambda \lambda_N v_u Z_{12} Z_{13}, \nonumber \\
C_{\tilde{\nu}_1 \tilde{\nu}_1 H_u} &\simeq &  - \sqrt{2} T_\nu Z_{11} Z_{12} - Y_\nu^2 v_u Z_{12} Z_{12},  \nonumber \\
C_{\tilde{\nu}_1 \tilde{\nu}_1 s} &\simeq & - 2 \kappa \lambda_N v_s  Z_{12} Z_{13} - \sqrt{2} T_{\lambda_N} Z_{12} Z_{13} - \lambda_N^2 v_s,  \label{approximation-1}
\end{eqnarray}
which indicate a hierarchical structure: $|C_{\tilde{\nu}_1 \tilde{\nu}_1 s}| \sim {\cal{O}}(100 {\rm GeV})$
and $|C_{\tilde{\nu}_1 \tilde{\nu}_1 H_d}|, |C_{\tilde{\nu}_1 \tilde{\nu}_1 H_u}|  \lesssim 10 {\rm GeV}$.
Furthermore, we consider two representative cases for the Higgs sector
\begin{itemize}
\item[I.]  $h_1$ corresponds to the SM-like Higgs boson, and $h_2$ and $h_3$ are decoupled from electroweak physics.

For this case, $S_{12} \simeq \sin \beta \sim 1$, $S_{11} \simeq \cos \beta$, ${a_u}_1 \simeq {a_d}_1$, and
\begin{eqnarray}
\sigma^{\rm SI}_{\tilde{\nu}_1-p} \propto \left ( \frac{\sqrt{2} T_\nu Z_{11} Z_{12} + Y_\nu^2 v_u Z_{12} Z_{12}}{(125 \rm GeV)^2 \ m_{\tilde{\nu}_1}} \right )^2. \label{h1-SM-like}
\end{eqnarray}
This formula indicates that the cross section is determined by $Y_\nu$ and $T_{\nu}\equiv Y_\nu A_\nu$, and may be suppressed if $\tilde{\nu}_1$
is $\tilde{x}$ dominated. We remind that a small $Y_\nu$ is not only favored by the recent XENON-1T constraints on
$\sigma^{\rm SI}_{\tilde{\nu}_1-p}$, but also consistent with the limitation on the non-unitarity of the $U$ matrix in neutrino sector.

As a comparison, one may also discuss the DM-nucleon scattering rate in the MSSM, which can be obtained from $\sigma^{\rm SI}_{\tilde{\nu}_1-p}$ by scaling the factor
$C_{\tilde{\nu}_1 \tilde{\nu}_1 h_1}^2/(4 C_{\tilde{\chi}_1^0 \tilde{\chi}_1^0 h_1}^2 m_{\tilde{\nu}_1}^2)$ as indicated by Eq.(\ref{Ratio}).
To be more specific, if $\tilde{\chi}_1^0$ is bino-dominated and meanwhile the higgsino mass $\mu$ is significantly larger than the bino mass $m_1$, we have \cite{Cao:2015efs}
\begin{eqnarray}
C_{\tilde{\chi}_1^0 \tilde{\chi}_1^0 h_1} \simeq \sqrt{4 \pi \alpha} \frac{m_Z}{\mu} (\sin 2 \beta + \frac{m_1}{\mu}).
\end{eqnarray}
Taking $m_{\tilde{\nu}_1} = m_{\tilde{\chi}_1^0} \simeq m_1 $, we conclude that the ratio is about $10 \times (Y_\nu Z_{12} \mu/m_1 )^4$
after neglecting unimportant terms.  This fact indicates that $\sigma^{\rm SI}_{\tilde{\nu}_1-p}$ in the ISS-NMSSM can be easily much lower than that in the MSSM.

\item[II.]  $h_1$ acts as the SM-like Higgs boson, $h_2$ is singlet dominated with $m_{h_2} \lesssim v$, and $h_3$ is decoupled.

In this case,  $S_{12} \simeq \sin \beta \sim 1$, $S_{11} \simeq \cos \beta$ and $S_{23} \sim 1$. At same time,
$|S_{13}|$ and $|S_{22}|$ are usually moderately larger than $|S_{21}|$, but all of them should be less than  about 0.1 to coincide
with the $125 {\rm GeV}$ Higgs data. Consequently, ${a_d}_1 \simeq {a_u}_1$,
${a_d}_2$ is much larger than ${a_u}_2$ since $\tan \beta \gg 1$, and ${a_d}_3, {a_u}_3 \simeq 0$ since they are suppressed by $1/m_{h_3}^2$.
$\sigma^{\rm SI}_{\tilde{\nu}_1-p}$ is then given by
\begin{eqnarray}
\sigma^{\rm SI}_{\tilde{\nu}_1-p} &\propto& \left [ \frac{C_{\tilde{\nu}_1 \tilde{\nu}_1 h_1}}{m_{h_1}^2 m_{\tilde{\nu}_1}} (\frac{S_{12}}{\sin\beta} + \frac{S_{11}}{\cos \beta} )
+ \frac{C_{\tilde{\nu}_1 \tilde{\nu}_1 h_2}}{m_{h_2}^2 m_{\tilde{\nu}_1}} (\frac{S_{22}}{\sin\beta} + \frac{S_{21}}{\cos \beta} ) \right ]^2 \nonumber  \\
&\propto& \left [ \frac{2 (C_{\tilde{\nu}_1 \tilde{\nu}_1 H_u} + S_{13} C_{\tilde{\nu}_1 \tilde{\nu}_1 s} ) }{ (125 \rm GeV)^2\  m_{\tilde{\nu}_1}}
+ \frac{C_{\tilde{\nu}_1 \tilde{\nu}_1 s}}{m_{h_2}^2 m_{\tilde{\nu}_1}} \frac{S_{21}}{\cos \beta}  \right ]^2,  \label{approximation-2}
\end{eqnarray}
where we used the approximation $C_{\tilde{\nu}_1 \tilde{\nu}_1 h_1} \simeq C_{\tilde{\nu}_1 \tilde{\nu}_1 H_u} + S_{13} C_{\tilde{\nu}_1 \tilde{\nu}_1 s}$ and
$C_{\tilde{\nu}_1 \tilde{\nu}_1 h_2} \simeq C_{\tilde{\nu}_1 \tilde{\nu}_1 s}$.

From above formulae, one can get following useful conclusions
\begin{itemize}
\item If $Y_\nu = 0$ and consequently $C_{\tilde{\nu}_1 \tilde{\nu}_1 H_u} \simeq 0$, we have
\begin{eqnarray}
C_{\tilde{\nu}_1 \tilde{\nu}_1 h_1} \simeq S_{13} C_{\tilde{\nu}_1 \tilde{\nu}_1 s} \simeq S_{13} C_{\tilde{\nu}_1 \tilde{\nu}_1 h_2} \ \ {\rm and} \ \
\frac{{a_d}_1}{{a_d}_2} \simeq \frac{S_{13} S_{11}}{S_{21}} \frac{m_{h_2}^2}{m_{h_1}^2} \sim {\cal{O}} (1). \label{relation}
\end{eqnarray}
For the typical case of ${a_d}_2 \simeq {a_d}_1$, Eq.(\ref{approximation-2}) is then reexpressed as
\begin{eqnarray}
\sigma^{\rm SI}_{\tilde{\nu}_1-p} \propto \left (  \frac{S_{13} C_{\tilde{\nu}_1 \tilde{\nu}_1 s}}{(125 \rm GeV)^2\  m_{\tilde{\nu}_1}} \right )^2,
\end{eqnarray}
which indicates that the magnitude of $\sigma^{\rm SI}_{\tilde{\nu}_1-p}$ is partially decided by the mixing $S_{13}$.
The implication of this special case is that the interaction of $\tilde{\nu}_1$ with the singlet fields alone can be responsible
for DM physics in the ISS-NMSSM, namely predicting correct relic density and also possibly sizable DM-nucleon scattering cross section.

We remind that Eq.(\ref{relation}) also holds if the element $S_{13}$ is not suppressed too much so that
$S_{13} C_{\tilde{\nu}_1 \tilde{\nu}_1 s} \gg C_{\tilde{\nu}_1 \tilde{\nu}_1 H_u}$.  We will study in detail this situation later.

\item In Eq.(\ref{approximation-2}), the first term comes from the interchange of $h_1$, and the second term denotes the contribution of $h_2$.
These two contributions are usually comparable in size since ${a_d}_1/{a_d}_2 \sim {\cal{O}}(1)$, and in some cases the latter may be more important.
We will show that the two contributions may interfere destructively or constructively in contributing to the cross section.
\end{itemize}

\end{itemize}

\begin{table}[t]
\begin{center}
\begin{tabular}{|c|c|c|c|c|c|}
  \hline
   parameter& value & parameter & value& parameter& value \\ \hline
  $\tan \beta $ & 15.8 & $\lambda$ & 0.22 & $\kappa$ & 0.17  \\
  $A_\lambda$ & 2150 $GeV$& $A_\kappa$ & -18 $GeV$& $\mu $ & 120.0 $GeV$ \\
  $m_{\tilde{q}}$ & 2000 $GeV$ & $m_{\tilde{l}}$ & 400 $GeV$ &$A_{u,c,d,s}$ & 2000 $GeV$ \\
  $A_{t,b}$ & -3000 $GeV$& $A_{e,\mu,\tau}$ &  400 $GeV$& $M_1$ & 400 $GeV$ \\
  $M_2$ & 800 $GeV$ & $M_3$ & 2400 $GeV$ & $m_{h_1}$ & 125.2 $GeV$ \\
  $m_{h_2}$ & 176.3 $GeV$ & $m_{h_3}$ & 2030 $GeV$ & $m_{A_1}$ & 67.7 $GeV$\\
  $m_{A_2}$ & 2030 $GeV$ & $m_{{\tilde{\chi}}_1^0}$& 106.9 $GeV$ &$m_{{\tilde{\chi}}_2^0}$ & 130.7 $GeV$\\
  $m_{{\tilde{\chi}}_3^0}$ & 189 $GeV$& $m_{{\tilde{\chi}}_1^{\pm}}$ & 121.9 $GeV$ & $m_{{\tilde{\chi}}_2^{\pm}}$ & 832 $GeV$\\
  $S_{11}$ & 0.064 & $S_{12}$ & 0.995& $S_{13}$ & 0.075\\
  $S_{21}$ & 0.015 & $S_{22}$ & 0.076& $S_{23}$ & 0.996\\
  $S_{31}$ & 0.997 & $S_{32}$ & 0.063& $S_{33}$ & 0.024\\
  \hline
\end{tabular}
\end {center}
\caption{Fixed parameters in the NMSSM sector when we present our numerical results. In this table, $h_1$ acts as the SM-like Higgs boson, $h_2$
and $A_1$ are singlet dominated scalars, the mass degenerate $h_3$ and $A_2$ correspond to the heavy Higgs bosons in the MSSM, and $S$ is the
rotation matrix to diagonalize the mass matrix for the CP-even Higgs bosons in the basis $(H_d, H_u, s)$. Note that since we have set the higgsino mass at
$120 {\rm GeV}$, which is motivated by naturalness argument, all the higgsino-dominated particles such as $\tilde{\chi}_1^0$, $\tilde{\chi}_2^0$ and
$\tilde{\chi}_1^\pm$ are light with mass around $120 {\rm GeV}$. Also note that the masses for the Higgs bosons are slightly altered by the parameters
in sneutrino sector through loop effects, so their values in this table are actually obtained for the case of $Y_\nu = \lambda_N=0$. }
\label{benchmark}
\end{table}

\section{Numerical Results}

In this section, we study the property of the sneutrino DM $\tilde{\nu}_1$ by presenting some numerical results. In order to illustrate the underlying physics as clearly as possible,
we first fix the parameters in the NMSSM sector, and give in Table \ref{benchmark} the values of some quantities which are relevant to our study. Then we adopt
the Metropolis-Hastings algorithm \footnote{To be more explicit, we adopt the likelihood function
${\cal{L}} = {\cal{L}}_{m_{h_1}} \times {\cal{L}}_{\Omega h^2} \times {\cal{L}}_{Br(B \to X_s \gamma)} \times  {\cal{L}}_{Br(B_s \to \mu^+ \mu^-)}$
for the Markov Chain Monte Carlo scan where ${\cal{L}}_{m_{h_1}}$, ${\cal{L}}_{\Omega h^2}$, ${\cal{L}}_{Br(B \to X_s \gamma)}$ and ${\cal{L}}_{Br(B_s \to \mu^+ \mu^-)}$ are likelihood functions for
experimentally measured SM-like Higgs boson mass, DM relic density, $Br(B \to X_s \gamma)$ and $Br(B_s \to \mu^+ \mu^-)$ respectively, which are taken to
be Gaussian distributed \cite{Likelihood,Cao:2016uwt}. } implemented in the code \textsf{EasyScan\_HEP} \cite{es}
to scan following parameter space in sneutrino sector \footnote{Since we concentrate on the property of $\tilde{\nu}_1$ instead of on neutrino oscillations, we set $\mu_X = 0$
for simplicity, and only consider the effects of the third generation sneutrinos by setting $Y_{\nu}=0$ and the diagonal elements of $m_{\nu}$ and $m_{x}$ at $1 {\rm TeV}$ for the other
two generation sneutrinos. With such a treatment, $\lambda_N$ in Eq.(\ref{sneutrino-scan}) actually corresponds to the (3,3)
element of the matrix $\lambda_N$ in Eq.(\ref{Superpotential}), and so are the parameters $Y_\nu$, $A_{\lambda_N}$, $A_\nu$, $m_\nu$ and $m_x$. }
\begin{eqnarray}\label{sneutrino-scan}
0 \leq \lambda_{N}, Y_{\nu}\leq 0.3, \quad  -1 {\rm TeV} \leq A_{\lambda_N}, A_\nu \leq 1 {\rm TeV},\quad 50{\rm ~GeV}\leq m_{\nu},  m_{x} \leq 150 {\rm ~GeV}.
\end{eqnarray}
In the calculation, we utilize the package \textsf{SARAH-4.11.0} \cite{sarah-1,sarah-2,sarah-3} to build the model and the code  \textsf{SPheno-4.0.3} \cite{spheno}
to generate the particle spectrum, and we consider following constraints
\begin{itemize}
\item $123 {\rm GeV} \leq m_{h_1} \leq  127 {\rm GeV}$, which is the most favored range of the SM-like Higgs boson mass by current LHC results \cite{Khachatryan:2016vau}. This constraint arises from the fact that the parameters in sneutrino sector can alter the Higgs boson mass spectrum through loop effects \cite{MSSM-ISS-10,MSSM-ISS-13,B-L-ISS-3}.
\item consistence of the Higgs properties with the data from LEP, Tevatron and LHC experiments. This is due to the consideration that the non-standard neutrinos may serve as the
decay products of the SM-like Higgs boson, and thus change the branching ratios of its decay into SM particles, and also the consideration that the moderately light $h_2$ may
induce sizable signals at the colliders. We implement the requirement by the packages \textsf{HiggsBounds-5.0.0} \cite{HiggsBounds} and \textsf{HiggsSignal-2.0.0} \cite{HiggsSignal}.
\item low energy flavor observables, such as $B \rightarrow X_s \gamma$, $B_s\rightarrow\mu^{+} \mu^-$ and $\Delta M_{B_s}$, and  muon anomalous magnetic momentum within $2\sigma$ range around its experimental central value. These observables can be calculated automatically by the code  \textsf{SPheno-4.0.3} under the instruction of the package \textsf{SARAH-4.11.0}.
\item $m_{\tilde{\nu}_1} < m_{\tilde{\chi}_1^0}$, and $0.107 < \Omega h^2 < 0.131$ in order to account for the Planck measurement of DM relic density at $2\sigma$ level \cite{Ade:2015xua}.
\end{itemize}

\begin{figure}[t]
\centering
\includegraphics[scale=0.4]{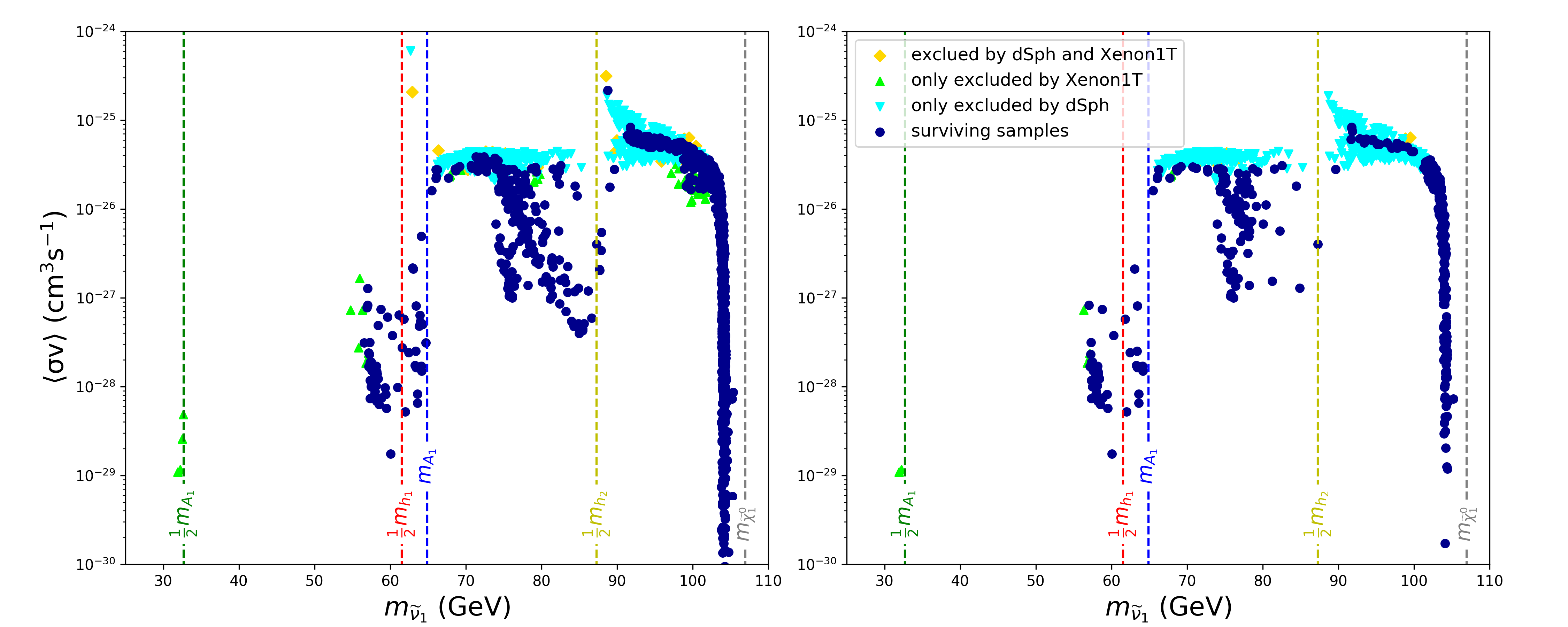}
\caption{{\bf Left panel:} Samples obtained from the scan with the constraints listed in the text considered, which are projected on
$\langle{\sigma}v\rangle_0-m_{\tilde{\nu}_1}$ plane.  We use the dark blue color to represent the samples that satisfy both the dSph constraint
and the XENON-1T constraint, and the lime color, the cyan color and the golden yellow color to denote those which are
exclude by either the XENON-1T constraint or the dSph constraint, or the both respectively.  Samples around the green, red and yellow
vertical lines annihilated in early Universe through the resonant $A_1$, $h_1$ and $h_2$ respectively, and those near the gray line
obtain the correct relic density mainly by the annihilation of the higgsinos. For the samples close from left to the blue line, the annihilation channel
$\tilde{\nu}_1 \tilde{\nu}_1 \to A_1 A_1 $ opens up in the early Universe, and soon becomes the dominant one with the
increase of $m_{\tilde{\nu}_1}$ up to about $100 {\rm GeV}$. {\bf Right panel:} similar to the left panel except that we further impose the constraint of
$Y_\nu v_u/(\lambda_N v_s) <0.1$ on the samples, which is motivated by the limitations from the non-unitary of neutrino
mixing matrix and the electroweak precision data.  \label{sigmaV}}
\end{figure}

\begin{figure}[t]
\centering
\includegraphics[scale=0.4]{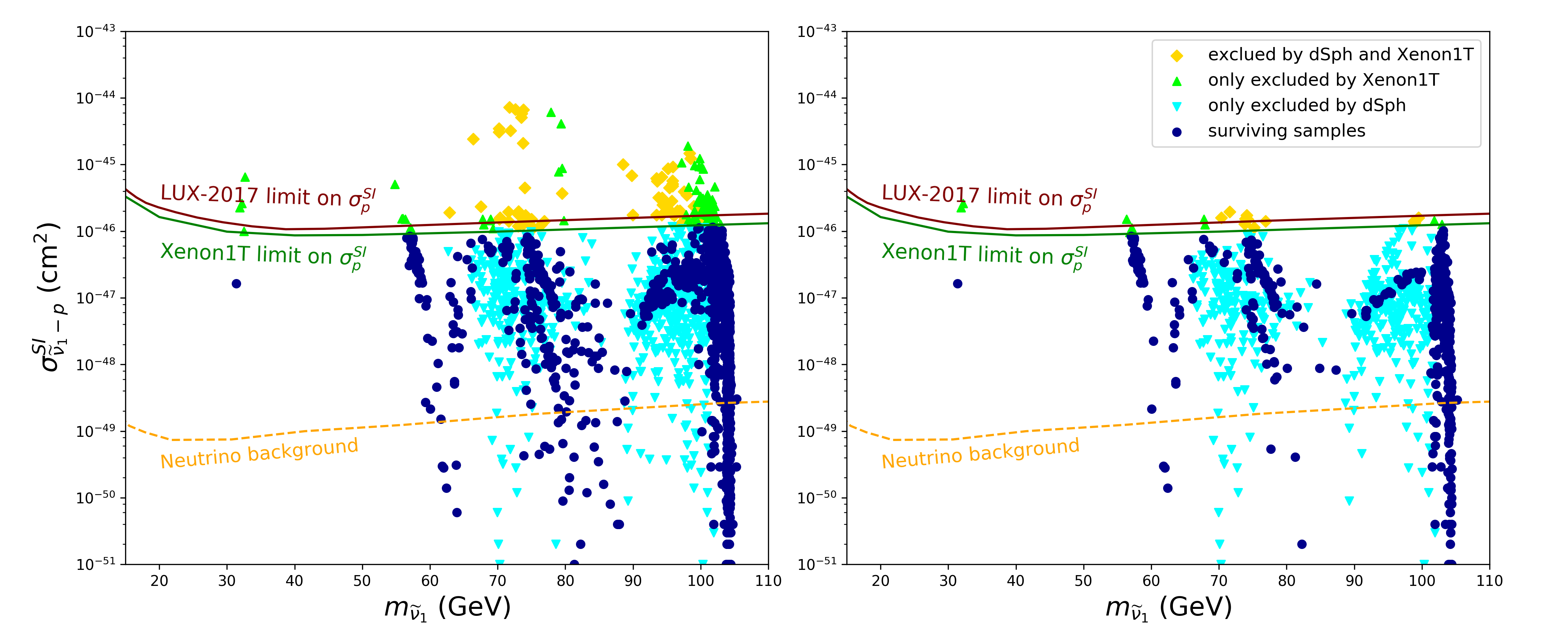}
\caption{Similar to the Fig. \ref{sigmaV}, but projected on $\sigma_{\tilde{\nu}_1-p}^{SI}-m_{\tilde{\nu}_1}$ plane.  \label{sigma_psi}}
\end{figure}

\begin{figure}[t]
\centering
\hspace{0.05cm} \includegraphics[height=7.2cm,width=7.5cm]{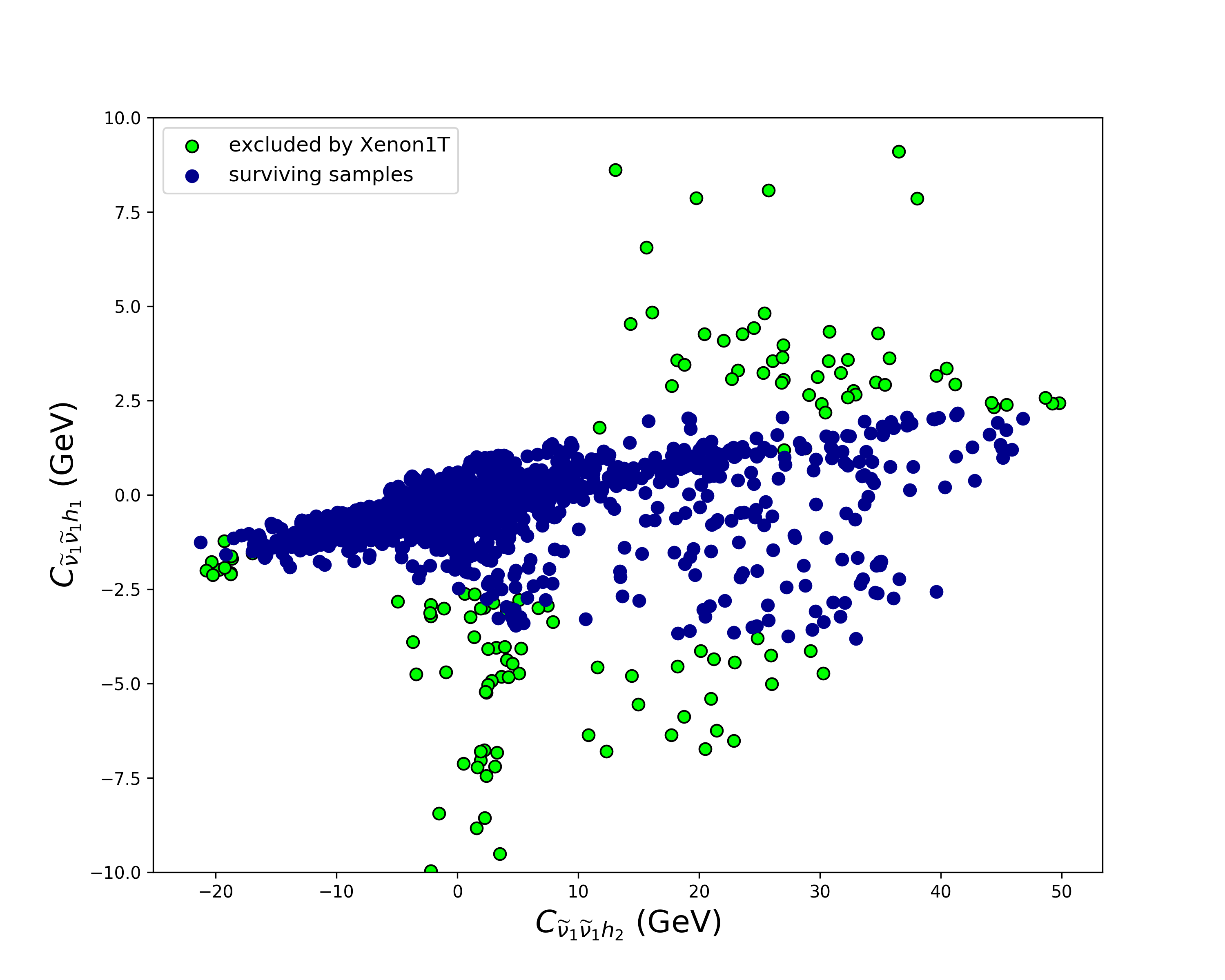} \hspace{-0.7cm}
\includegraphics[height=7.2cm,width=7.5cm]{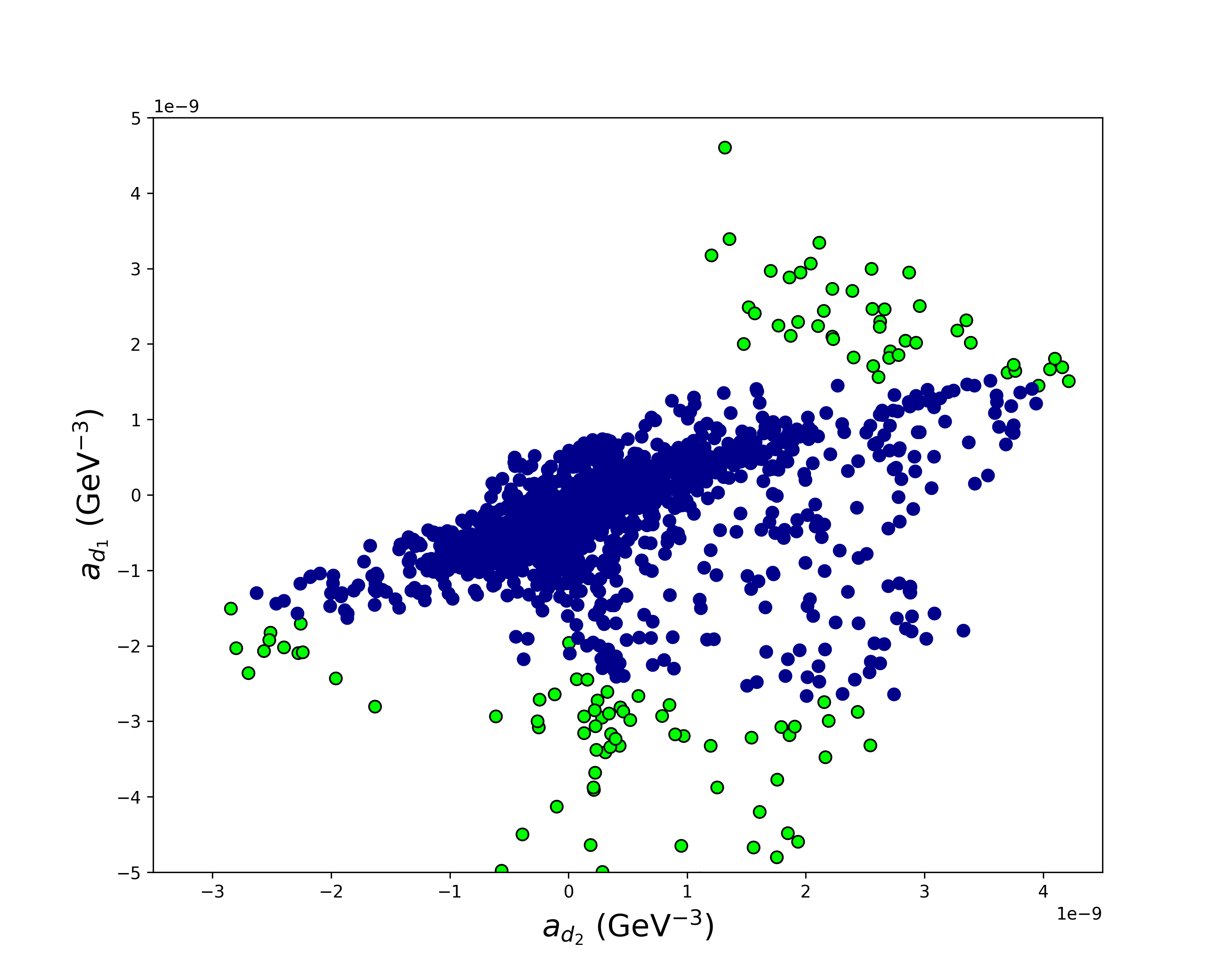}
\vspace{-0.2cm}
\caption{Same samples as those in the left panel of Fig. \ref{sigma_psi}, but projected on the $C_{\tilde{\nu}_1 \tilde{\nu}_1 h_1}-C_{\tilde{\nu}_1 \tilde{\nu}_1 h_2}$
plane (left panel) and the ${a_d}_1-{a_d}_2$ plane (right panel). These samples are classified by whether they are excluded by the XENON-1T experiment
(marked by green color) or not (dark blue color).  \label{couplings}}
\end{figure}

As we introduced in last section, the sneutrino sector of the ISS-NMSSM  provides great flexibility to account for DM physics. In our study, we consider about three thousand
samples obtained from the scan with the constraints considered. We find $|Z_{11}| < 0.1$ for all the samples, and  $|Z_{12}| \simeq |Z_{13}| \simeq 1/\sqrt{2}$ for
a sizable portion of the samples. In Fig.\ref{sigmaV}, we project the samples on $\langle{\sigma}v\rangle_0-m_{\tilde{\nu}_1}$ plane with the dark blue color to represent
those that satisfy both the dSph constraint and the XENON-1T constraint, and the lime color, the cyan color and the golden yellow color to denote those which are
excluded by either the XENON-1T constraint or the dSph constraint, or the both respectively.  In implementing the dSph constraint, we use the data provided by Fermi-LAT collaboration
\cite{Fermi-Web}, and adopt the likelihood function proposed in \cite{Likelihood-dSph,Zhou}, while in imposing the XENON-1T constraint, we use directly the 90\%
exclusion limits on the SI DM-nucleon scattering cross-section of the recent XENON-1T experiment \cite{Aprile:2017iyp}.

From Fig.\ref{sigmaV}, one can infer that for the samples around the green, red and yellow vertical lines, $\tilde{\nu}_1$ annihilated in early Universe mainly by the
resonant Higgs mediated processes $\tilde{\nu}_1 \tilde{\nu}_1^R \to A_1^\ast \to f \bar{f}$,
$\tilde{\nu}_1 \tilde{\nu}_1 \to h_1^\ast \to f \bar{f}$ and  $\tilde{\nu}_1 \tilde{\nu}_1 \to h_2^\ast \to A_1 A_1, f \bar{f}, W^+ W^-$ respectively, and for the samples
near the gray line, it achieves right relic density mainly by the annihilation of the higgsinos.  Moreover, the annihilation channel
$\tilde{\nu}_1 \tilde{\nu}_1 \to A_1 A_1 $ opens up in the early Universe for the samples close from left to the blue line, and it soon becomes
the dominant one with the
increase of $m_{\tilde{\nu}_1}$ up to about $100 {\rm GeV}$. Since this annihilation is a $s$-wave dominant process, the dSph constraint is rather strong
to exclude a large portion of the samples. While on the other hand, there still exist various ways to escape the constraint as we introduced in last section.
Different from the left panel in Fig.\ref{sigmaV} where only the constraints listed in the text are considered, the right panel further considers
the constraint $Y_\nu v_u/(\lambda_N v_s) <0.1$ on the samples.
This constraint is motivated by the limitations on the non-unitary of neutrino mixing matrix \cite{MSSM-ISS-10} and the
electroweak precision data \cite{MSSM-ISS-13}. We remind that the masses of $A_1$, $h_1$ and $h_2$ are slightly altered by the radiative correction from the sneutrino sector,
and the positions of the vertical lines only act as a rough indication of the masses.

Next we consider the SI DM-nucleon scattering rate, which is the focus of this work. In Fig.\ref{sigma_psi}, we project the samples of Fig. \ref{sigmaV} on
$\sigma_{{\tilde{\nu}_1}-p}^{SI}-m_{\tilde{\nu}_1}$ plane with the same color convention as that of Fig. \ref{sigmaV}. As expected in last section, the constraint
from the XENON-1T experiment is rather weak
on the sneutrino DM in the ISS-NMSSM, and only a small portion of the samples are excluded. Especially if we further require $Y_\nu v_u/(\lambda_N v_s) <0.1$, only few
samples are excluded. We emphasize that in the ISS-NMSSM, the SI cross section can be lower than the neutrino background even for light higgsinos, and consequently the DM
may never be probed in DD experiments.

In the following, we present more information about the SI cross section. In Fig.\ref{couplings}, we only consider the samples in the left panel of Fig. \ref{sigma_psi},
and project them on $C_{\tilde{\nu}_1 \tilde{\nu}_1 h_1}-C_{\tilde{\nu}_1 \tilde{\nu}_1 h_2}$ plane (left panel) and ${a_d}_1-{a_d}_2$ plane (right panel) respectively, where the green samples
are excluded by the XENON-1T experiment, and the dark blue ones are not. The left panel indicates that  $|C_{\tilde{\nu}_1 \tilde{\nu}_1 h_1}| \lesssim 4 {\rm GeV}$,
and $C_{\tilde{\nu}_1 \tilde{\nu}_1 h_2}$ varies in a much wider range from $-20 {\rm GeV}$ to $50 {\rm GeV}$  for the surviving samples. It also indicates  that the couplings
$C_{\tilde{\nu}_1 \tilde{\nu}_1 h_1}$ and $C_{\tilde{\nu}_1 \tilde{\nu}_1 h_2}$ seem to be roughly linear dependent for most samples.  The underlying reason for the correlation
is that $C_{\tilde{\nu}_1 \tilde{\nu}_1 h_2} \simeq C_{\tilde{\nu}_1 \tilde{\nu}_1 s}$ and $C_{\tilde{\nu}_1 \tilde{\nu}_1 h_1} \simeq C_{\tilde{\nu}_1 \tilde{\nu}_1 H_u} + S_{13} C_{\tilde{\nu}_1 \tilde{\nu}_1 s} \simeq S_{13} C_{\tilde{\nu}_1 \tilde{\nu}_1 s}$ where we used the fact $ |C_{\tilde{\nu}_1 \tilde{\nu}_1 s}| \gg |C_{\tilde{\nu}_1 \tilde{\nu}_1 H_u}|$
(for similar discussions, see Eq. \ref{relation}).  The right panel shows that $-2 {\rm GeV^{-3}} \lesssim {a_d}_1  \lesssim 1 {\rm GeV^{-3}}$
and $-3 {\rm GeV^{-3}} \lesssim {a_d}_2  \lesssim 1.5 {\rm GeV^{-3}}$  for the surviving samples, and a similar correlation between ${a_d}_1$ and
${a_d}_2$ exists for most samples. About Fig.\ref{couplings} three points should be noted. First, for the typical setting of the NMSSM parameters in Table \ref{benchmark},
the coefficients $a_i$ obey the relations: ${a_u}_1 \simeq {a_d}_1$,  $|{a_d}_2|$ is several times larger than $|{a_u}_2| $ due to the large $\tan \beta$,
and $|{a_d}_1| \gg  |{a_u}_3|, |{a_d}_3|$. As for ${a_d}_1$ and ${a_d}_2$, their magnitudes may be comparable, and they can interfere constructively or destructively
in contributing to the cross section. Second, since $a_{d2} \varpropto C_{\tilde{\nu}_1 \tilde{\nu}_1 h_2}/m_{h_2}^2 $, the range of $C_{\tilde{\nu}_1 \tilde{\nu}_1 h_2}$
must be narrowed correspondingly to survive the XENON-1T constraints if we choose a lighter $h_2$. In this case, more parameter space of the ISS-NMSSM will be limited by the DD experiments.
Third, in case of $C_{\tilde{\nu}_1 \tilde{\nu}_1 H_u} \simeq 0$ where the correlation holds, only the interactions of $\tilde{\nu}_1$ with the singlet Higgs field are significant.
These interactions alone can be responsible for the right relic density, and meanwhile contribute to the cross section. This cross section, however,
is usually lower than the bound of the XENON-1T experiment, and is thus experimentally favored.

\begin{figure}[t]
\centering
\includegraphics[scale=0.35]{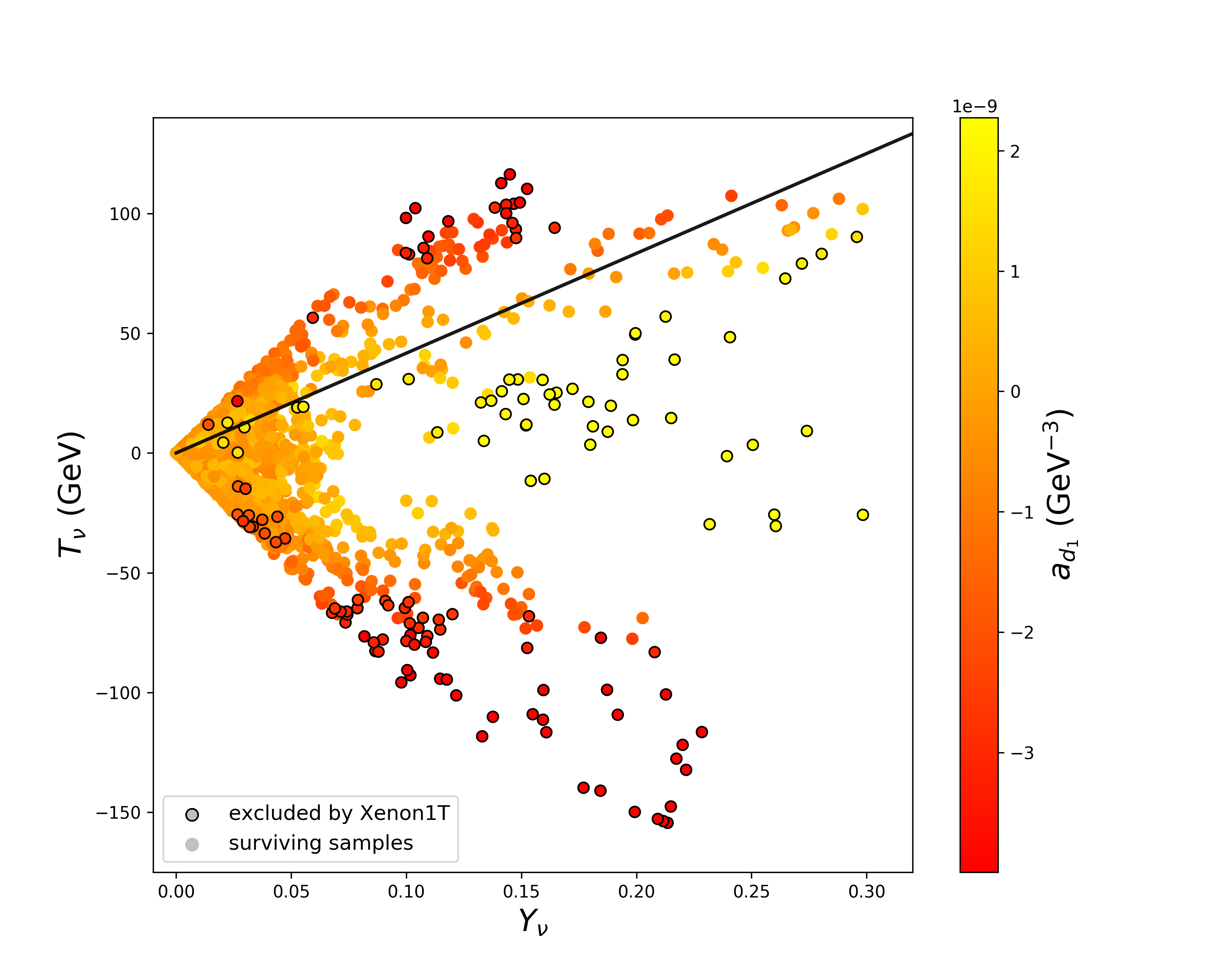}
\caption{Similar to Fig.\ref{couplings}, but projected on $T_{\nu}-Y_{\nu}$ plane with $T_{\nu} \equiv Y_\nu A_\nu$. The circled samples are excluded by the XENON-1T experiment,
and the colors represent different values of $a_{d_1}$, which are defined by the colored bar on the right side of the figure.  \label{Tv_Yv}}
\end{figure}

\begin{figure}[t]
\centering
\includegraphics[scale=0.35]{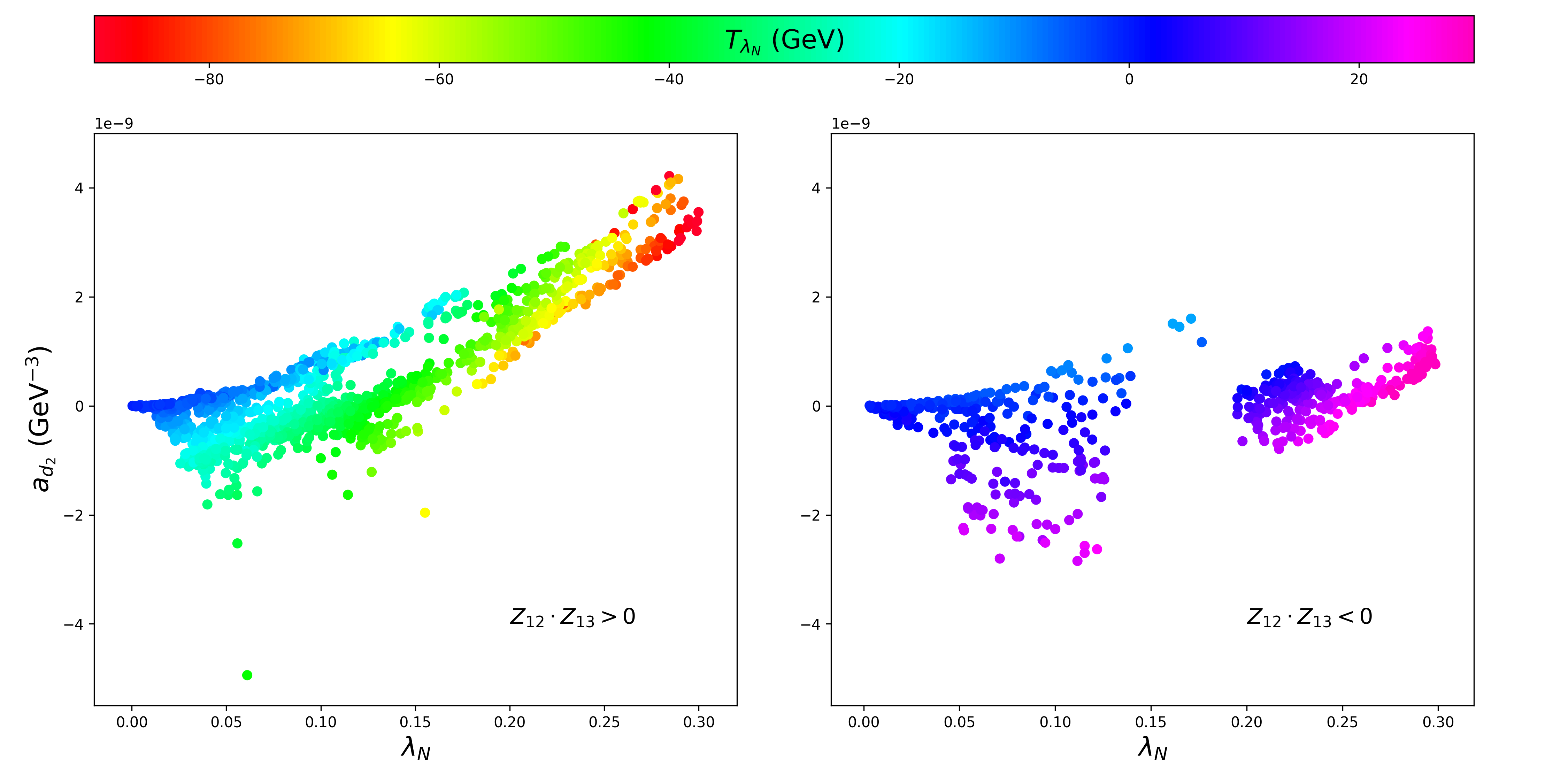}
\caption{Similar to Fig.\ref{couplings}, but projected on $a_{d_2}-{\lambda_N}$ plane with $Z_{12} Z_{13}>0$ case (left panel) and $Z_{12} Z_{13}<0$ case (right panel) respectively.
The colors represent the values of $T_{\lambda_N} \equiv \lambda_N A_{\lambda_N} $, which are shown on top of the figure.    \label{ad2_lambdaN}}
\end{figure}

Finally, we consider the dependence of the cross section on the parameters in sneutrino sector. In Fig.\ref{Tv_Yv}, we project the samples
in Fig.\ref{couplings} on $T_{\nu}-Y_{\nu}$ plane, where the colors correspond to the values of $a_{d1}$ and the circled samples are excluded
by the XENON-1T experiment.  This figure indicates that the sample with a large $Y_\nu$ and/or a large $T_\nu$ tends to predict large
$|{a_d}_1|$ and $\sigma_{\tilde{\nu}_1-p}$, and the parameter region preferred by the experiment is $Y_\nu \lesssim 0.15$ and
$|T_\nu | \lesssim 100 {\rm GeV}$. This fact can be understood from Eq.(\ref{h1-SM-like}) by noting that the $h_2$ contribution is insensitive to
the two parameters.  We note that for the samples along the black line direction, their predictions on $|{a_d}_1|$ are usually small even for large
$Y_\nu$ and $T_\nu$. We checked that it is due to the cancelation between the $Y_\nu$ contribution and
the $T_\nu$ contribution. We also note that there exist samples which correspond to small $Y_\nu$ and $T_\nu$, but are excluded by the XENON-1T
experiment. We checked that these samples correspond to a quite large $|{a_d}_2|$ with ${a_d}_2/{a_d}_1 > 0$.

In Fig.\ref{ad2_lambdaN}, we project the samples in Fig.\ref{couplings} on $a_{d_2}-{\lambda_N}$ plane with the colors indicating the values of $T_{{\lambda}_N}$.
The left panel and the right panel correspond to $Z_{12} Z_{13}>0$ case and $Z_{12} Z_{13}< 0$ case respectively. This figure indicates that for $Z_{12} Z_{13}>0$ case,
$T_{\lambda_N}$ prefers to be negative, while for $Z_{12} Z_{13}< 0$ case, it tends to positive. In any case, the effect of $T_{\lambda_N}$ is to cancel the
$\lambda_N$ contribution to ${a_d}_2$. This can be understood by the formula
\begin{eqnarray}
{a_{d}}_2 &=& -\frac{g}{8 m_W} \frac{C_{\tilde{\nu}_1 \tilde{\nu}_1 h_2}}{m_{h_2}^2 m_{\tilde{\nu}_1}} \frac{S_{21}}{\cos \beta},\nonumber\\
&\simeq &  \frac{g}{8 m_W} \frac{S_{21}}{\cos \beta} \frac{2 \kappa \lambda_N v_s  Z_{12} Z_{13} + \sqrt{2} T_{\lambda_N} Z_{12} Z_{13} + \lambda_N^2 v_s}{m_{h_2}^2 m_{\tilde{\nu}_1}},
\end{eqnarray}
where we used the approximation $C_{\tilde{\nu}_1 \tilde{\nu}_1 h_2} \simeq C_{\tilde{\nu}_1 \tilde{\nu}_1 s}$ and Eq.(\ref{approximation-1}). We remind that it is due to
the cancelation, $\lambda_N$ as large as 0.3 is still allowed by the XENON-1T experiment. We also remind that the allowed values of $\lambda_N$ and $T_{\lambda_N}$ by
the XENON-1T experiment depend on our choice of $m_{h_2}$.

\section{LHC constraints on the model}

\begin{figure}[t]
\centering
\includegraphics[height=7.2cm,width=10cm]{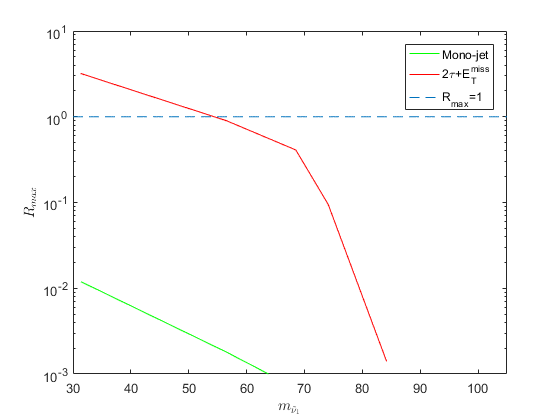}
\vspace{-0.4cm}
\caption{$R_{max}$ as a function of $m_{\tilde{\nu}_1}$ for the Mono-jet signal (green line) and $2 \tau + E_{T}^{miss}$ signal (red line).
Note that in this work, the higgsino mass is fixed at $120 {\rm GeV}$, and consequently $m_{\tilde{\chi}^\pm} = 121.9 {\rm GeV}$,
$\sigma_{\rm 8TeV}(p p \to \tilde{\chi}_1^\pm \tilde{\chi}_1^\mp )\approx 0.728 {\rm pb}$ and
$\sigma_{\rm 13 TeV}(p p \to \tilde{\chi}_1^\pm \tilde{\chi}_1^\mp )\approx 1.46 {\rm pb}$, where the cross sections are calculated
at next-to-leading order by the code Prospino \cite{prospino}.} \label{simulation-results}
\end{figure}

In this section, we examine the constraints from the direct searches for electroweakinos at the LHC on the samples
considered in last section. Since $Br(\tilde{\chi}_{1,2}^0 \to \tilde{\nu}_1 \nu_\tau ) =
Br(\tilde{\chi}_{1,2}^0 \to \tilde{\nu}_1^R \nu_\tau ) \simeq 50\%$,  $Br(\tilde{\chi}_{1}^\pm \to \tilde{\nu}_1
\tau^\pm ) = Br(\tilde{\chi}_{1}^\pm \to \tilde{\nu}_1^R \tau^\pm ) \simeq 50\%$ for the samples and $\tilde{\nu}_1^R$ is
long-lived at colliders due to its nearly degeneracy with $\tilde{\nu}_1$ in mass, we consider
the Mono-jet signal from the processes $p p \to  \tilde{\chi}_1^\pm \tilde{\chi}_{1,2}^0, \tilde{\chi}_{1,2}^0 \tilde{\chi}_{1,2}^0 j$
and the $2 \tau + E_T^{miss}$ signal from the process  $p p \to \tilde{\chi}_1^\pm  \tilde{\chi}_1^\mp$ in our discussion.

For the signal of Mono-jet$ + E_T^{miss}$, we consider the analyses at 8-TeV LHC by ATLAS and CMS collaborations \cite{monojet-1,monojet-2,monojet-3} and
13-TeV LHC  by ATLAS collaboration \cite{monojet-4}, all of which have been encoded in the package  CheckMATE \cite{cmate-1,cmate-2,cmate-3}.
The common requirements of the analyses are: (1) an energetic jet with $p_T>100 {\rm GeV}$ and possible existence of
one additional softer jet with $\Delta\phi(j_1, j_2)<2.5$ to suppress large QCD dijet background; (2) large missing energy, typically $E_T^{miss} > 150 {\rm GeV}$;
(3) vetoing any event with isolated leptons. With regard to the signal of two hadronic $\tau$s plus $E_T^{miss}$,
the strongest limit comes from the analyses of the direct Chargino/Neutralino production at the 8-TeV LHC by ATLAS and
CMS collaborations \cite{tausLHC1, tausLHC2} and the 13-TeV LHC by ATLAS collaboration \cite{tausLHC3}. As far as our case
(i.e. fixed $\mu$ at $120 {\rm GeV}$) is concerned,
the analysis in \cite{tausLHC1} imposes stronger constraint than that in \cite{tausLHC3}, which can be learned from
figure 7 in \cite{tausLHC3}. The underlying reason is that the analysis in \cite{tausLHC3} focuses on
heavy Chargino case, which requires more energetic jets and larger missing energy than the former.
Moreover, we note that the constraint of the analysis in \cite{tausLHC2} is similar to that in \cite{tausLHC1} for
$m_{\tilde{\chi}_1^\pm} < 200 {\rm GeV}$, which can be learned by comparing   figure 5 in \cite{tausLHC2} with figure 7
in \cite{tausLHC1}. So in this work, we only consider the analysis in \cite{tausLHC1} on the $2 \tau + E_T^{miss}$
signal. We implement this analysis in the package CheckMATE with the corresponding validation presented in appendix.

\begin{table}[t]
  \centering
  \begin{tabular}{|c|c|c|c|}
    \hline
    $m_{\tilde{\nu}_1} ({\rm GeV})$ & $SR^{*}$ & $\epsilon $ & R \\ \hline
    31.3 & SR-C1C1 & 0.19 & 3.2 \\ \hline
    56.5 & SR-C1C1 & 0.06 & 0.90 \\ \hline
    68.5 & SR-C1C1 & 0.03 & 0.41 \\ \hline
    74.1 & SR-C1C1 & 0.014 & 0.095 \\ \hline
    84.2 & SR-DS-lowMass & 0.006 & 0.0014 \\ \hline
    105.3 & - & 0 & 0 \\ \hline
  \end{tabular}
  \caption{Detailed information about the analysis of $2 \tau + E_T^{miss}$ signal in \cite{tausLHC1} for six parameter points.
    $SR^{*}$ stands for the SR with the largest expected sensitivity, and $\epsilon$ is the net cut efficiency of the signal events.}
  \label{tab-lhc}
\end{table}

To study these signals, we first use the package SARAH \cite{sarah-1,sarah-2,sarah-3} to generate the model files
of the ISS-NMSSM in UFO format \cite{ufo}. Then we use the simulation tools MadGraph/MadEvent \cite{mad-1,mad-2} to generate
the parton level events of the processes with Pythia6  \cite{pythia} for parton fragmentation and hadronization, and
Delphes \cite{delphes} for the fast simulation of the ATLAS or CMS detector. Finally we use the improved CheckMATE to implement
the cut selections of the analyses.

For each of the analyses, we consider the signal region (SR) with the largest expected sensitivity
for a given $m_{\tilde{\nu}_1}$ \footnote{For each experimental analysis, the expected sensitivity of the
$i$-th SR is defined as $R_{\rm EXP,i} = S_i/S_{95,i}^{\rm EXP}$ where $S_i \equiv S^0_i - 1.96 \times \Delta S$ with
$S^0_i$ denoting the number of signal events after cuts and $\Delta S$ being its statistical uncertainty, and
$S_{95,i}^{\rm EXP}$ stands for expected limit at 95\% confidence level for same SR. The most sensitive SR corresponds
to $R_{\rm EXP} = Max (R_{\rm EXP,i})$. }, and calculate its $R$ value defined by $R \equiv S/S_{95}^{OBS}$, where $S$ stands
for the number of signal events in the SR with the statistical uncertainty considered and $S_{95}^{OBS}$ denotes the observed limit
at 95\% confidence level for the SR. For the signal which corresponds to several experimental analyses,
we select the largest $R$ among the analyses, denoted by $R_{max}$ hereafter, to parameterize the capability of the LHC
in exploring the parameter point. If $R_{max}$ is larger than unity, the point is excluded and otherwise it is allowed.
In Fig.\ref{simulation-results}, we present our results of $R_{max}$ for the Mono-jet signal
(green line) and the $2\tau + E^T_{miss}$ signal (red line) respectively. This figure indicates that the constraints from
the Mono-jet signal is very weak, and $R_{max}$ reaches only $0.01$ in best case. The underlying reason is that the
experimental analyses require a relatively large $E_T^{miss}$, which can not be satisfied for most of the events.
By contrast, $R_{max}$ for the $2\tau + E^T_{miss}$
signal increases monotonously with the enlarged mass splitting between $\tilde{\chi}_1^\pm$ and $\tilde{\nu}_1$, and
for $m_{\tilde{\nu}_1} \lesssim 55 {\rm GeV}$ it exceeds unity. In order to understand the features of the analysis
on the $2 \tau + E^T_{miss}$ signal, we choose six parameter points and provide in Table \ref{tab-lhc}  more information about
the analysis in \cite{tausLHC1}. As can be seen from this table, the cut efficiency is quite large for $m_{\tilde{\nu}_1} \simeq 30 {\rm GeV}$,
reaching about $19\%$, and it drops quickly with the increase of $m_{\tilde{\nu}_1}$ to $0.6\%$ for  $m_{\tilde{\nu}_1} \simeq 84 {\rm GeV}$.

\section{Phenomenology of the ISS-NMSSM}

In the ISS-NMSSM, the DM candidate may be the lightest sneutrino or the lightest neutralino. In this section, we only briefly sum up the phenomenology of the
former case. Some of our viewpoints may be applied to the latter case, which will enrich the well studied phenomenology of the NMSSM.

In the ISS-NMSSM, the impact of the sneutrino DM on the phenomenology is reflected in following aspects:
\begin{itemize}
\item Relaxing greatly the parameter space of the NMSSM, and meanwhile maintaining the naturalness of the model. As we pointed out in \cite{Cao:2016cnv}, so far
the DD experiments have put very strong constraints on the natural NMSSM, and consequently it is not easy to get parameter points coinciding
with the constraints from DD experiments. In the ISS-NMSSM, however, $\tilde{\nu}_1$ can serve as a viable DM candidate if
$ {\rm Min}(m_{h_1}, m_{A_1}) < 2 m_{\tilde{\chi}_1^0}$ or if the lightest higgsino corresponds to
$\tilde{\chi}_1^0$, which have been illustrated before. These conditions can be easily satisfied in the ISS-NMSSM, and consequently new features in
comparison with the original  NMSSM may appear in Higgs physics as well as in sparticle physics. Taking the parameter point in Table \ref{benchmark} as an example,
we found that it can not get the proper relic density and meanwhile predict an unacceptable large SD cross section for DM-nucleon scattering
in the framework of the NMSSM \cite{Cao:2016nix}. In the ISS-NMSSM, however, it becomes a phenomenologically viable point.
\item Existence of relatively light particles, such as non-standard neutrinos and light Higgs bosons, which, beside
exhibiting themselves at colliders by exotic signals,
may serve as the decay products of the Higgs bosons and sparticles.  This feature makes the search for new particles at colliders quite complicated. For examples,
we find from the samples that the non-standard neutrinos may be as light as $30 {\rm GeV}$. In this case, a left-handed slepton may decay dominantly into one of the neutrinos plus
a higgsino-dominated neutralino by the neutrino Yukawa interaction. As a result, the signature of the slepton is distinct from that in the NMSSM. Moreover,
since the neutrino has a small left-handed neutrino component, it may be produced in association with one active neutrino at the LHC \cite{ISS-NMSSM-5}, or with one lepton \cite{Das:2014jxa}, or in pairs \cite{Neutrino-Signal-1,Neutrino-Signal-2,Neutrino-Signal-3}.
Obviously, how to detect these signals is an open question.

\item Existence of new interactions which alter the properties of the particles in the NMSSM, and also induce new contribution to some observables. For example, the neutrino Yukawa
interaction in the ISS-NMSSM can not only change the decay modes of the higgsinos and the left-handed sleptons in the NMSSM, but also contribute to Higgs boson masses
\cite{MSSM-ISS-10,MSSM-ISS-13,B-L-ISS-3}, lepton flavor violating processes \cite{MSSM-ISS-1,MSSM-ISS-3,MSSM-ISS-4,MSSM-ISS-6,MSSM-ISS-14,MSSM-ISS-16,MSSM-ISS-18}
as well as muon anomalous magnetic momentum \cite{B-L-ISS-8,B-L-ISS-10}.
\end{itemize}
Due to these aspects, the phenomenology of the ISS-NMSSM is quite rich, and may be different from that of the NMSSM.

As far as sparticles are concerned, the speciality of the ISS-NMSSM comes from the fact that the couplings of $\tilde{\nu}_1$ with the other particles are usually
suppressed, and meanwhile it carries a certain lepton flavor number if $Y_\nu$ is flavor diagonal.  As a result, heavy sparticles will not decay directly into $\tilde{\nu}_1$,
but instead they first decay into a relatively light sparticle with stronger couplings \cite{MSSM-ISS-16,Banerjee:2016uyt,Chakrabarty:2014cxa}. This lengthened decay chain makes the decay
products of the parent sparticle quite model dependent. For example, if the sleptons are lighter than the higgsinos, the signature of the higgsinos usually corresponds
to multi-lepton final state \cite{Arina:2015uea,Arina:2013zca}, which is different from the
final states discussed in last section. We remind that in principle
$Y_\nu$ may be flavor non-diagonal, and consequently $\tilde{\nu}_1$ will not have a definite lepton flavor number any more.
This further complicates the sparticle decays.

\section{Conclusions}

Given the increasing tension between naturalness and the DD experiments for customary neutralino DM candidate in supersymmetric theories, we discuss the feasibility that
the lightest sneutrino acts as a DM candidate to alleviate the tension. For this end, we assume certain symmetries, and extend the field content of
the NMSSM in an economical way to incorporate the inverse seesaw mechanism into the framework for neutrino mass. We point out that the resulting
theory called ISS-NMSSM not only inherits all the merits of the NMSSM and the seesaw mechanism, but also exhibits new features in both DM physics and
sparticle phenomenology. Especially by choosing the sneutrino as DM candidate, we find by analytic formulae that the DM-nucleon scattering
rate is usually suppressed in comparison with the neutralino DM in the MSSM, and consequently the constraints from the DD experiments are no
longer strong. We also find that the interactions of the sneutrino with the singlet Higgs field alone can account for the measured relic density,
and meanwhile predict acceptable cross sections for both direct and indirect DM search experiments.
We show these features numerically in physical parameter space, which is obtained by fixing the parameters
in the NMSSM sector and scanning the parameters in the sneutrino sector with various experimental constraints
(including the LHC search for $2 \tau + E_T^{miss}$ and {\rm Mono-jet}$+E_T^{miss}$ signals) considered.
Finally, we also briefly discuss the phenomenology of the ISS-NMSSM, and point out that it is quite rich and distinct from that of the NMSSM.
Given that the LHC experiments have not probed any signals of sparticles, the ISS-NMSSM may deserve a comprehensive study in near future.

Before we end this work, we'd like to compare briefly the ISS-NMSSM with the Type-I seesaw extension of the NMSSM proposed in \cite{NMSSM-SS-1}.
In the Type-I seesaw extension, only right-handed neutrino fields are introduced to generate neutrino mass, and the corresponding neutrino Yukawa
couplings are of ${\cal{O}}(10^{-6})$, which is at same order as the electron Yukawa coupling in the SM,  given that the masses for the right-handed
neutrinos are about $1 {\rm TeV}$. In both models, the singlet Higgs field plays an important role in various aspects, including generating the
higgsino mass and the heavy neutrino masses dynamically, mediating the transition between $\tilde{\nu}_1$ pair and higgsino pair to keep
them in thermal bath in early Universe, acting as DM annihilation final state or mediating DM annihilations, as well as affecting DM-nucleon scattering rate.
Consequently  both models can yield in certain parameter space thermal DM and a sneutrino-nucleon scattering cross section
compatible with DD limits of the recent XENON-1T experiment. On the other hand, the essential difference of the two models comes from following
two aspects. One is that in order to accommodate the experimental data for the neutrino oscillations, the electroweak
precision measurements and the lepton-flavor violations, one can choose in the ISS-NMSSM a flavor-blind neutrino
Yukawa couplings by encoding all the flavor structures into the small lepton-number violating parameter $\mu_X$ as indicated in Eq.(\ref{mu-X}). This will make
the non-unitary limitation mentioned in Eq.(\ref{non-unitarity}) easily satisfied. By contrast, there is no such freedom in the type-I seesaw extension, and one has to rely on the
neutrino Yukawa couplings to account for all the experimental data. So we conclude that the ISS-NMSSM provides more theoretical flexibility in
accommodating the data and at same time much richer phenomenology at colliders \cite{Neutrino-Signal-3}. The other different comes
from the signature of the heavy neutrinos \cite{Das:2015toa}. In the Type-I seesaw extension, due to the
Majorana nature of the heavy neutrinos, its associated production with one lepton at the LHC usually results in same-sign di-lepton signal, while in
the ISS-NMSSM due to the pseudo-Dirac nature of the neutrinos, the process usually leads to tri-lepton signals. A more dedicated
comparison of the two models will be carried out in our forthcoming work.

\section*{Acknowledgement}

 We thank Prof. Xiaojun Bi and Yufeng Zhou for helpful discussion about dark matter indirect detection experiments. This work is supported by the National Natural Science Foundation of China (NNSFC) under grant No. 11575053.

\appendix
\section{Appendix}

In this section, we validate our code for all SRs in \cite{tausLHC1}. We work in the MSSM, and consider four cases
which correspond to $\tilde{\chi}_1^{\pm} \tilde{\chi}_1^{\mp}$ and $\tilde{\chi}_1^{\pm} \tilde{\chi}_2^{0}$ productions
with both $\tilde{\chi}_1^\pm$ and $\tilde{\chi}_2^0$ being wino-dominated,
$\tilde{\chi}_1^{\pm} \tilde{\chi}_1^{\mp}$ production with $\tilde{\chi}_1^\pm$  being wino-dominated,
$\tilde{\tau}_R \tilde{\tau}_R$ production and $\tilde{\tau}_L \tilde{\tau}_L$
production respectively. For each validation, we generate 10000 events in the way introduced in Section 4. Our results are presented
in Table \ref{case1}, \ref{case2}, \ref{case3} and \ref{case4} respectively.  These tables indicate that we can reproduce the
ATLAS results for case 1-3 at $20\%$ level, and case 4 at $30\%$.

\begin{table}[h!]
\centering
\begin{tabular}{|c|c|c|c|c|c|}
\hline
\multicolumn{2}{|c|}{\multirow{2}{*}{$(m_{\tilde{\chi}_1^{\pm}, \tilde{\chi}_2^0}, m_{\tilde{\chi}_1^0}, m_{\tilde{\tau},\tilde{\nu}})$ {[}GeV{]}}} & ATLAS & \multicolumn{3}{c|}{CheckMATE}                                                            \\ \cline{3-6}
\multicolumn{2}{|c|}{}        & $R_{\rm ATLAS}$ &  $SR^{*}$ &   $R$ & Diff [\%] \\ \hline
P1                              & 300,100,200                              & 1.0     &  SR-C1N2    &  0.90    &  -10.0    \\ \hline
P2                              & 200,75,137.5                             & 1.0     &  SR-C1N2    &   1.06    &  6.0    \\ \hline
\end{tabular}%
\caption{Validation of the $\tilde{\chi}_1^{\pm} \tilde{\chi}_1^{\mp}$ and $\tilde{\chi}_1^{\pm} \tilde{\chi}_2^{0}$ production processes
at the 8-TeV LHC by assuming $m_{\tilde{\chi}_1^\pm} = m_{\tilde{\chi}_2^0}$ and $m_{\tilde{\tau}} = m_{\tilde{\nu}} =
(m_{\tilde{\chi}_1^\pm} + m_{\tilde{\chi}_2^0})/2$. $R_{\rm ATLAS}$ in the table is the result obtained by ATLAS collaboration,
which is taken from the exclusion line of Fig.7a in \cite{tausLHC1}.
$SR^{*}$ and $R$ have same meanings as those in Table \ref{tab-lhc}, and ${\rm Diff} \equiv (R - R_{\rm ATLAS})/R_{\rm ATLAS}$,
which parameterizes the deviation of our calculation from its corresponding ATLAS result.}
\label{case1}
\end{table}

\begin{table}[h!]
\centering
\begin{tabular}{|c|c|c|c|c|c|}
\hline
\multicolumn{2}{|c|}{\multirow{2}{*}{$(m_{\tilde{\chi}_1^{\pm}},m_{\tilde{\chi}_1^0},m_{\tilde{\nu},{\tilde{\tau}}})$ {[}GeV{]}}} & ATLAS & \multicolumn{3}{c|}{CheckMATE}                                                            \\ \cline{3-6}
\multicolumn{2}{|c|}{}                 & $R_{\rm ATLAS}$     &  $SR^{*}$ & $R$ & Diff [\%]        \\ \hline
P1                & 300,80,190         & 1.0   &  SR-DS-highMass      &  0.81    &    -19.0  \\ \hline
P2                & 200,75,137.5       & 1.0   &  SR-DS-highMass      &   0.96    &     -4.0  \\ \hline
\end{tabular}%
\caption{Similar to Table \ref{case1}, but for the $\tilde{\chi}_1^{\pm} \tilde{\chi}_1^{\mp}$ production
process with the corresponding ATLAS results plotted in Fig.7b of \cite{tausLHC1}.}
\label{case2}
\end{table}

\begin{table}[h!]
\centering
\begin{tabular}{|c|c|c|c|c|c|}
\hline
\multicolumn{2}{|c|}{\multirow{2}{*}{$(m_{{\tau}_R},m_{\tilde{\chi}_1^0})$ {[}GeV{]}}} & ATLAS & \multicolumn{3}{c|}{CheckMATE}                                                            \\ \cline{3-6}
\multicolumn{2}{|c|}{}         & $R_{\rm ATLAS}$     & $SR^{*}$ & $R $ & Diff [\%] \\ \hline
P1            & 300,100        & 1.0   &  SR-DS-highMass   &    0.96    & -4.0    \\ \hline
P2            & 200,100        & 1.0   &  SR-DS-highMass   &    0.86    & -14.0    \\ \hline
P3            & 150,100        & 1.0   &  SR-DS-lowMass    &    1.18    & 18.0     \\ \hline
\end{tabular}%
\caption{Similar to Table \ref{case1}, but for the  $\tilde{\tau}_R \tilde{\tau}_R$ production process
with the corresponding ATLAS results plotted in Fig.8a of \cite{tausLHC1}. }
\label{case3}
\end{table}

\begin{table}[h!]
\centering
\begin{tabular}{|c|c|c|c|c|c|}
\hline
\multicolumn{2}{|c|}{\multirow{2}{*}{$(m_{{\tau}_L},m_{\tilde{\chi}_1^0})$ {[}GeV{]}}} & ATLAS & \multicolumn{3}{c|}{CheckMATE}                                                            \\ \cline{3-6}
\multicolumn{2}{|c|}{}      & $R_{\rm ATLAS}$     &  $SR^{*}$ & $ R$ & Diff [\%]  \\ \hline
P1    & 300,100        & 1.0     &  SR-DS-highMass    &   1.15   &  15.0    \\ \hline
P2    & 200,100        & 1.0     &  SR-C1C1           &   1.27   &  27.0    \\ \hline
P3    & 150,100        & 1.0     &  SR-DS-lowMass     &   1.11   &  11.0     \\ \hline
\end{tabular}%
\caption{Similar to Table \ref{case1}, but for the $\tilde{\tau}_L \tilde{\tau}_L$ production process with
the corresponding ATLAS results plotted in Fig.8b of \cite{tausLHC1}. }
\label{case4}
\end{table}

\end{document}